\documentclass[structabstract]{aa}  
\usepackage{graphicx}
\usepackage{longtable,lscape}
\usepackage{txfonts}
\usepackage{natbib}
\bibpunct{(}{)}{;}{a}{}{,} 
\def\diff{\mathrm d}

\begin{document}
   \title{Type Ia supernovae and the $^{12}$C+$^{12}$C reaction rate}
\authorrunning{E. Bravo et al.}
\titlerunning{SNIa and the $^{12}$C+$^{12}$C rate}

   \author{E. Bravo\inst{\ref{inst1}}
	   \and
          L. Piersanti\inst{\ref{inst2},\ref{inst3}}
           \and
          I. Dom\'\i nguez\inst{\ref{inst4}}
           \and
          O. Straniero\inst{\ref{inst2},\ref{inst3}}
           \and
          J. Isern\inst{\ref{inst5},\ref{inst6}}
           \and
          J.A. Escartin\inst{\ref{inst7}}
           }

   \institute{Dept. F\'\i sica i Enginyeria Nuclear, Univ. Polit\`ecnica de
              Catalunya, Carrer Pere Serra 1-15, 08173 Sant Cugal del Vall\`es, Spain\\   
              \email{eduardo.bravo@upc.edu} \label{inst1}
         \and
              INAF - Osservatorio Astronomico di Teramo,
              via mentore Maggini snc, 
              64100 Teramo, Italy\\
              \email{[piersanti;straniero]@oa-teramo.inaf.it} \label{inst2}
	 \and
	      INFN Sezione di Napoli, 
	      I–80126 Napoli, Italy \label{inst3}
	 \and
              Departamento de F\'\i sica Te\'orica y del Cosmos, 
              Universidad de Granada, 
              18071 Granada, Spain\\
              \email{inma@ugr.es} \label{inst4}
         \and
              Institut de Ci\`encies de l'Espai (CSIC), Campus UAB, 
              08193 Bellaterra, Spain\\
              \email{isern@aliga.ieec.uab.es} \label{inst5}
         \and
	      Institut d’Estudis Espacials de Catalunya, 
              Ed. Nexus-201, c/Gran Capita 2-4, 
              08034 Barcelona, Spain \label{inst6}
	 \and
	      Dept. F\'\i sica i Enginyeria Nuclear, Univ. Polit\`ecnica de
              Catalunya, Carrer Comte d'Urgell 187, 08036 Barcelona, Spain\\   
              \email{jose.antonio.escartin@upc.edu} \label{inst7}
	      }

   \date{Received ; accepted }

 
  \abstract
{Even if the $^{12}$C+$^{12}$C reaction plays a central role in the ignition of Type Ia
supernovae (SNIa), the experimental determination of its cross-section at astrophysically 
relevant energies ($E\lesssim2$~MeV) has never been made.
The profusion of resonances throughout the measured energy range has led to
speculation that there is an unknown resonance at $E_0\sim1.5$~MeV
possibly as strong as the one measured for the resonance at 2.14~MeV, i.e.
$\left(\omega\gamma\right)_\mathrm{R}=0.13$~meV.}
{We study the implications that
such a resonance would have for our knowledge of the physics of SNIa, paying
special attention to the phases that go from the crossing of the ignition curve
to the dynamical event.}
{We use one-dimensional hydrostatic and hydrodynamic codes to follow the evolution of accreting
white dwarfs until they grow close to the Chandrasekhar mass and explode as SNIa. 
In our simulations, we account for a low-energy resonance by exploring the parameter space 
allowed by experimental data.}
{A change in the $^{12}$C+$^{12}$C rate similar to the one explored here would have profound
consequences for
the physical conditions in the SNIa explosion, namely the central density, neutronization, thermal profile,
mass of the convective core, location of the runaway hot spot, or time elapsed since crossing
the ignition curve. For instance, with the largest resonance strength we use, the
time elapsed since crossing the ignition curve to the supernova event is shorter by a factor ten
than for models using the standard rate of $^{12}$C+$^{12}$C, and the runaway
temperature is reduced from $\sim8.14\times10^8$~K to $\sim4.26\times10^8$~K. 
On the other hand, a resonance at 1.5~MeV, with a strength ten
thousand times \emph{smaller} than the one measured at 2.14~MeV, but with an $\alpha$/p
yield ratio substantially different from 1 would have a sizeable impact on the degree of
neutronization of matter during
carbon simmering.
}
{A robust understanding of the links between SNIa properties and their progenitors will not be
attained until the $^{12}$C+$^{12}$C reaction rate is measured at energies $\sim1.5$~MeV.}

   \keywords{nuclear reactions, nucleosynthesis, abundances --
	     supernovae: general --
             white dwarfs   
               }

   \maketitle
%

\section{Introduction}

The idea that Type Ia supernovae (SNIa) are triggered by explosive carbon burning in degenerate
material dates back to the seminal paper of \citet{hoy60}: ``Pure carbon is explosive through
$^{12}\mathrm{C}(^{12}\mathrm{C},\alpha)^{20}\mathrm{Ne}$ and
$^{12}\mathrm{C}(^{12}\mathrm{C},\mathrm{p})^{23}\mathrm{Na}$ at a temperature somewhat
less than \mbox{$1.5\times10^9$}~degrees, even on a timescale as short as 1 second''.
Since that time, the paradigm of SNIa has changed from carbon detonation
\citep{arn69}, first, to carbon deflagration \citep{nom76}, and, finally, to the currently favoured
paradigm: a delayed detonation starting as a carbon deflagration \citep[][delayed-detonation
model]{kho91}. Shortly after the formulation of the delayed-detonation paradigm,
\citet{hoe95} and \citet{hk96} showed that this kind of model can account for the basic
observational properties (brightness, light curves) of normal and subluminous SNIa. 
Nowadays, sofisticated three-dimensional simulations of SNIa based on different flavours of the
delayed-detonation model \citep[e.g.][]{gam03,mea09,bra09b,rop11} are confronted
with data in order to explain increasingly subtle observational details \citep{maz07,kas09,mae10b}.
These
numerical simulations have taught us that the outcome of the explosion is very sensitive to the
runaway conditions \citep[e.g.][]{sei11}, which are set during the so-called carbon
simmering phase, the
pre-explosive period that begins when carbon starts burning at a slow pace and ends when the
nuclear timescale becomes comparable to the white dwarf (WD) sound crossing time,
$t_\mathrm{sound}\simeq0.1$~s \citep{pir08}.
The evolution of the progenitor WD during the simmering phase (which is difficult to address
because it is on the borderline between hydrostatic and hydrodynamic phenomena) is controlled by the
competition between cooling processes (neutrino emission and convection) and heating
by a dominant nuclear reaction: the fusion of two $^{12}$C nuclei at temperatures in the range
$2\times10^8~\mathrm{K}$-- $10^9$~K. 

The experimental measurement of the cross-section of the $^{12}\mathrm{C}+^{12}\mathrm{C}$
reaction (carbon fusion reaction) at low energies has advanced slowly in the past few decades. At the
time of formulation of
the carbon detonation model for SNIa \citep{arn69}, the latest known experimental results 
extended down to a
center of mass energy $E_\mathrm{cm}=3.23$~MeV \citep{pat69}. Seven years later, when the
deflagration model was proposed, the reference $^{12}\mathrm{C}+^{12}\mathrm{C}$ reaction rate
evaluated by  \citet[][hereafter FCZ75]{fcz75} was based on experimental measurements at
$E_\mathrm{cm}>2.45$~MeV \citep{maz73}. When \citet{kho91} formulated the delayed detonation
paradigm, the calculation of the carbon fusion rate was based on the \citet[][hereafter
CF88]{cf88}
fit to experimental cross-sections down to the same energy $E_\mathrm{cm}$ ($\sim2.5$~MeV), although with
higher resolution \citep[e.g.][]{bec81}. The rate from CF88 differs from that of FCZ75
by less than 20\% at $T\gtrsim10^9$~K, and by less than a factor 2.5 at $T\gtrsim10^8$~K. 
In the past 30 years, the lower limit to $E_\mathrm{cm}$ reached experimentally has decreased to
only 2.10~MeV \citep{agu06,bar06,spi07}, because of experimental difficulties in reducing the
background
related to secondary reactions induced by hydrogen and deuterium contamination in the carbon targets.
Present-day data exhibit pronounced resonances throughout the measured energy range. 
\citet{spi07}, indeed, found a strong resonance at $E_\mathrm{R}=2.14$~MeV, close to the low-energy limit of
their measurements.
It turns out that the Gamow energies relevant to the carbon fusion reaction during the simmering
process lie far below the lower energy tested experimentally. For instance, for temperatures in the
range $2\times10^8~\mathrm{K}$-- $10^9$~K, the Gamow energies go from $E_\mathrm{G}\pm\Delta
E_\mathrm{G} = 0.82\pm0.14$~MeV to $E_\mathrm{G}\pm\Delta E_\mathrm{G} = 2.4\pm0.5$~MeV. 
Even though the most recent low-energy measurements give an average astrophysical factor $\sim2$-4
times smaller than the value recommended by CF88, the cross-sections measured at
$E_\mathrm{cm}\lesssim3$~MeV are rather uncertain, which explains why the 
$^{12}\mathrm{C}+^{12}\mathrm{C}$ reaction rate formulated by CF88
is still nowadays the reference rate for astrophysical applications. 

There has been lively discussion about the extrapolation of the
$^{12}\mathrm{C}+^{12}\mathrm{C}$
rate throughout the Gamow energy range. On the one hand,
theoretical phenomenological models predicting the behavior of the non-resonant 
astrophysical factor at low energies, based on
different approaches, differ by up to $\sim2$-3 orders of magnitude at
$E_\mathrm{cm}\sim1$-1.5~MeV \citep{gas05,agu06,jia07}.
Most of these predictions give rates well below the {\sl standard} rate of CF88 \citep[see, e.g.,
Fig. 3 in][]{str08}. On the other hand, the profusion of measured resonances has led to speculation
about the hypothetical presence of a resonance within the Gamow energy range,
say at $E_\mathrm{cm}\sim1.5$~MeV, that could completely dominate the reaction rate at the
densities and temperatures characteristic of the carbon simmering phase of SNIa. 

Our aim is to explore the consequences that a resonance in the
$^{12}\mathrm{C}+^{12}\mathrm{C}$ reaction near $E_\mathrm{cm}\sim1.5$~MeV would have for the
physics of SNIa. As a strong resonance at this energy is expected to dominate the rate, we can
disregard the uncertainties in the behavior of the non-resonant part of the astrophysical factor
at low energies, and adopt for it the CF88 rate. 
This kind of speculation about a hypothetical low-energy resonance in the carbon fusion reaction
rate has been addressed in several works. \citet{coo09b} studied the changes induced by
such a resonance
on the physics of superburst ignition on accreting neutron star crusts. 
\citet{ben10} and \citet{ben10b} explored the impact on the
s-process induced by 
an increase in the carbon fusion reaction rate up to a factor 50,000 at
$T=5\times10^8$~K.
Finally, \citet{iap10}
analyzed the influence of uncertainties in the $^{12}\mathrm{C}+^{12}\mathrm{C}$ reaction at
$T\lesssim5\times10^8$~K on the ignition of the WD core. They found that the ignition density
depends slightly on the presence of a low-energy resonance while the ignition temperature is almost
unaffected. However, their results are influenced by the small value of the resonance
strength they assumed, which implied that their reaction rate exceeded that from CF88
by only a factor of two. 

As there exists no theoretical framework that allows us to predict the location and strength
of the resonances in the $^{12}\mathrm{C}+^{12}\mathrm{C}$ reaction, the best approach to
studying their consequences is an exploration of the parameter space of a low-energy resonance.
In the next section, we determine the resonance energy and strength that are compatible with the
available cross-section data, and discuss the expected implications for our understanding of
SNIa. In Section~\ref{wdevol}, we present the results of our simulations of the evolution of a WD
from mass accretion to supernova explosion. These
simulations have been obtained with the use of one-dimensional hydrostatic, hydrodynamic, and
nucleosynthetic codes where the $^{12}\mathrm{C}+^{12}\mathrm{C}$ reaction rate was computed
consistently assuming different resonance properties. 
To cover a wider range of evolutionary scenarios (in particular, a wider range of
accretion rates
that would lead to different density-temperature tracks), we adopt in Section~\ref{runaway} a
somewhat different approach: we follow the evolution of a homogeneous region through the latest
stages of accretion until the dynamical instability with different physico-chemical parameters. In
this section, we study the dependence of the global neutronization of the convective core of
the WD during carbon simmering, and the runaway temperature of convective bubbles. Finally, we
present our conclusions in Section~\ref{theend}.

\section{Low-energy resonance}\label{ler}

We explore the consequences of a resonance in the neighborhood of
$E_0=1.5\pm0.2$~MeV with a strength limited by the measured cross-sections at low energy. 
The low-energy measurements that have provided the tightest constraints to date are those obtained by 
\citet[][hereafter S07]{spi07}, who explored energies $E>2.10$~MeV and found that the resonance
structure continues down to this energy limit. They also found a resonance at
$E_\mathrm{cm}=2.14$~MeV, although this resonance has a quite limited impact on carbon burning: it
speeds up slightly the carbon fusion rate at temperatures in the range
$T\sim6\times10^8$-$1.2\times10^9$~K (e.g., by a factor $1.8$ at $T=8\times10^8$~K).
On the other hand, the $\alpha$-channel yield of this resonance is larger
than the p-channel by a factor $\sim5$, at variance with the {\sl standard} formulation of CF88,
in which both channels have similar strength.

Our $^{12}\mathrm{C}+^{12}\mathrm{C}$ reaction rate is given by the sum of
the nonresonant contribution (that we calculate following  CF88) plus a resonant
one that accounts for the resonance found by S07 at $E_\mathrm{cm}=2.14$~MeV, and the assumed
low-energy {\sl ghost resonance} (LER). We compute the contribution of both resonances to the
carbon
fusion reaction rate as (see, for instance, S07):
\begin{eqnarray}
\langle\sigma
v\rangle^\mathrm{R} & = & 2.28\times10^{-24}T_9^{-3/2}\bigg[\exp\left(-24.8/T_9\right)
\nonumber \\
& & + 
\frac{\left(\omega\gamma\right)_\mathrm{R}}{0.13~\mathrm{meV}}\exp\left(-11.6E_\mathrm{R}
/T_9\right)\bigg]
~\mathrm{cm}^3\mathrm{s}^{-1}\,,
\label{eqres}
\end{eqnarray}
\noindent where $E_\mathrm{R}$ is the energy, in MeV, at which a resonance is assumed, and
$\left(\omega\gamma\right)_\mathrm{R}$ is the {\sl ghost resonance} strength.

Although there exist in the literature some discussions of the properties of such a resonance
that are compatible with current experimental data, different authors allow for different limits.
Hence, we repeat here the derivation of the maximum possible strength of a LER
\citep[e.g.][]{coo09b}. 

As a first approach, we demand that the {\sl ghost resonance} at $E_\mathrm{R}$ contributes to the
cross-section at $E_\mathrm{cm}=2.10$~MeV less than 10\% of the value 
measured by S07 at the same energy,
$\sigma_\mathrm{exp}(2.1~\mathrm{MeV})<0.8$~nb. Assuming a narrow resonance, as
are all known resonances of the carbon fusion reaction, $\Gamma_\mathrm{R}\approx40$-100~keV
\citep{agu06}, the energy dependent cross-section is given by the Breit-Wigner formula
\citep[e.g.][]{clay68,rol88}:
\begin{equation}
 \sigma_\mathrm{R}\left(E\right) = \frac{0.6566}{\hat{A}E\left(\mathrm{MeV}\right)}
\frac{\left(\omega\gamma\right)_\mathrm{R}
\Gamma_\mathrm{R}}{\left(E-E_\mathrm{R}\right)^2+\Gamma_\mathrm{R}^2/4}~\mathrm{b}\,,
\label{eqbw}
\end{equation}
\noindent where $\hat{A}=6$ is the reduced baryon number.
Substituting $E=2.10$~MeV, $\Gamma_\mathrm{R}=100$~keV, and $\sigma_\mathrm{R}<0.08$~nb we get
\begin{eqnarray}
 \left(\omega\gamma\right)_\mathrm{R} & < & 7.3\times10^{-10} E\left(\mathrm{MeV}\right)
\frac{\left(E-E_\mathrm{R}\right)^2+\Gamma_\mathrm{R}^2/4}{\Gamma_\mathrm{R}} \nonumber\\ 
 & = & 
15.4\left[\left(2.10-E_\mathrm{R}\right)^2+2.5\times10^{-3}\right]~\mathrm{meV}\,,
\end{eqnarray}
\noindent where $E_\mathrm{R}$ is in MeV. Thus, the resonance strength is limited to be less than 9.9,
5.6, and 2.5~meV for $E_\mathrm{R}=1.3$, 1.5, and 1.7~MeV, respectively. We note that a smaller value
of the resonance width, $\Gamma_\mathrm{R}<100$~keV, would result in still higher upper limits to
the resonance strength.

A second, similar, approach to the determination of the maximum possible strength of the {\sl ghost
resonance} makes use of the recommended value of the modified astrophysical factor, $S^\ast$,
defined as
\begin{equation}
 S^\ast\left(E\right)=\sigma\left(E\right) E \exp\left(87.21E^{-1/2}+0.46E\right)\,,
\end{equation}
\noindent where $E$ is in MeV. The average experimental $S^\ast$ at the lowest energies is
$\sim0.7\times10^{16}$~MeV~b \citep{str08}, while the value recommended by CF88 is
$3\times10^{16}$~MeV~b. Taking an approximate value of the experimental modified astrophysical
factor $S_\mathrm{exp}^\ast\left(2.10~\mathrm{MeV}\right)\lesssim10^{16}$~MeV~b, and using
Eq.~\ref{eqbw} for the energy-dependent cross-section due to the {\sl ghost resonance}, we obtain
\begin{eqnarray}
 \left(\omega\gamma\right)_\mathrm{R} & \lesssim & \frac{\hat{A}}{0.6566~\mathrm{MeV~barn}}
\frac{\left(E-E_\mathrm{R}\right)^2+\Gamma_\mathrm{R}^2/4}{\Gamma_\mathrm{R}}
S_\mathrm{exp}^\ast\left(E\right) \nonumber\\
 & & \times\exp\left(-87.21E^{-1/2}-0.46E\right)\,,
\end{eqnarray}
\noindent which, after substituting $E=2.10$~MeV and $\Gamma_\mathrm{R}=100$~keV, gives
\begin{equation}
 \left(\omega\gamma\right)_\mathrm{R} \lesssim
2.54\left[\left(2.10-E_\mathrm{R}\right)^2+2.5\times10^{-3}\right]~\mathrm{meV}\,,
\end{equation}
\noindent again with $E_\mathrm{R}$ in MeV. In this case, the resonance strength is limited to 1.6,
0.92, and 0.41~meV for $E_\mathrm{R}=1.3$, 1.5, and 1.7~MeV, respectively. These limits increase
by a factor $\sim10$ if the resonance width is $\Gamma_\mathrm{R}=10$~keV.

A quite different approach, found in the literature, consists in \emph{assuming} that the
modified astrophysical factor \emph{remains constant} at energies below 2~MeV. This assumption
leads to much lower resonance strengths at the resonance energies we explore. However, we note
that these are \emph{not upper limits} to the resonance strength compatible with experimental data.

\begin{figure}[tb]
\centering
  \resizebox{\hsize}{!}{\includegraphics{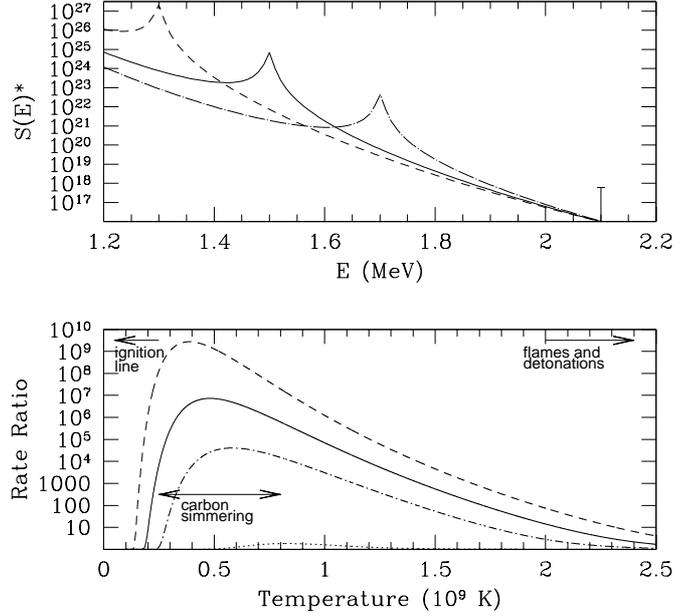}}
\caption{Astrophysical factor and rate of the $^{12}$C+$^{12}$C reaction due to a resonance
at an energy $E_\mathrm{R}=1.3$~MeV (dashed line), 1.5~MeV (solid line), and 1.7~MeV (dot-dashed
line), with resonance strengths of 16 meV, 9.2 meV, and 4.1 meV, respectively {\bf Top.} Modified
astrophysical factor as a function of energy. The bar at the bottom right shows the maximal
astrophysical factor measured by S07 at an energy of $\sim2.1$~MeV ($\sigma<0.8$~nb),
which is the most constraining low energy measurement to date. The quoted strengths of the
resonances would be compatible with a $\sigma$ two orders of magnitude smaller than that measured
by S07 (if the resonance width is $\Gamma_\mathrm{R}\approx10$~keV). {\bf
Bottom}. Ratio of our $^{12}$C+$^{12}$C reaction, accounting for a LER, to the
CF88 reaction rate, as a function of temperature. The dotted line close to the horizontal axis is
the ratio of the rate accounting for the resonance found by S07 to the CF88 rate. The arrows
mark the temperature ranges relevant to carbon ignition, carbon simmering, and the explosive
burning in a subsonic flame or a detonation. The gap between carbon simmering and flames reflects
the dominance of thermal conduction and shock heating over nuclear energy release in stationary
explosive burning fronts. 
}
\label{fig1}
\end{figure}

In Fig.~\ref{fig1} (top panel), we show the modified astrophysical factor of the $^{12}$C+$^{12}$C
reaction as a function of energy assuming a strong LER, compatible with the
limits
derived above, $S^\ast\left(2.10~\mathrm{MeV}\right)\lesssim10^{16}$~MeV~b. As might be
expected, the farther the energy of
the assumed resonance is from the measured energies, the larger is the maximum resonance strength
that is compatible with the data. We can see that a shift of $-0.2$~MeV in $E_\mathrm{R}$
allows us to increase the peak of the astrophysical factor by approximately two orders of magnitude.
The bottom panel of Fig.~\ref{fig1} shows the ratio of our $^{12}$C+$^{12}$C rate, which we calculated
as the sum of the CF88 rate and the resonant rate given by Eq.~\ref{eqres}, to the non-resonant
rate (CF88), as a function of temperature. The assumption of a strong \emph{ghost resonance} of the
aforementioned characteristics leads to a substantial enhancement of the carbon fusion rate in
the temperature range $2\times10^8$-- $2\times10^9$~K. While this temperature range depends
scarcely on the
assumed $E_\mathrm{R}$ (the peak of the rate ratio shifts from $3.9\times10^8$~K to
$5.8\times10^8$~K when the resonance energy changes from $E_\mathrm{R}=1.3$~MeV to 1.7~MeV), the
maximum ratio changes by as much as four orders of magnitude between the extreme explored values of
$E_\mathrm{R}$. 

\begin{figure}[tb]
\centering
  \resizebox{\hsize}{!}{\includegraphics{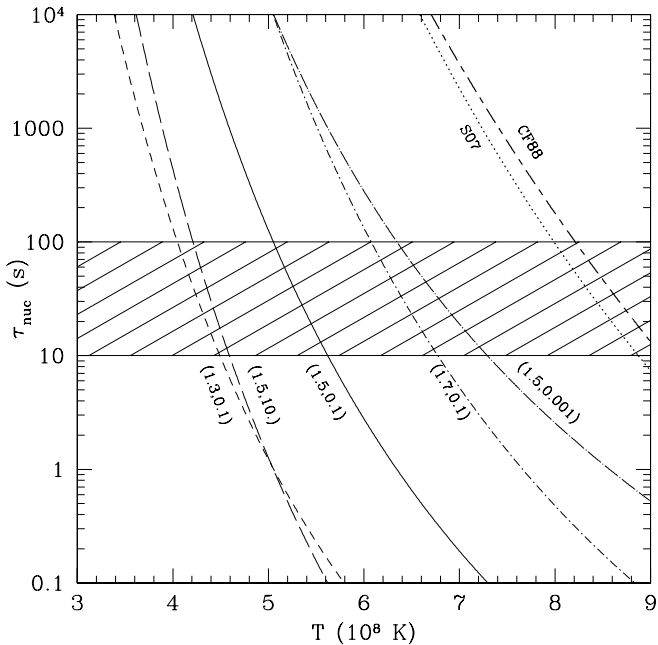}}
\caption{Nuclear timescale vs. temperature. The nuclear timescale has been computed as
$\tau_\mathrm{nuc}=\left(2\rho N_\mathrm{A} Y_{12} <\sigma v>\right)^{-1}$, where $N_\mathrm{A}$
is Avogadro's number, $Y_{12}=0.5/12$~mol~g$^{-1}$ is the molar fraction of $^{12}$C, and $<\sigma
v>$ is the rate of the $^{12}$C+$^{12}$C reaction. There are shown curves belonging to the CF88
non-resonant rate, the non-resonant plus the resonance at 2.1~MeV measured by S07, and
the non-resonant plus the 2.1~MeV resonance plus a LER. For the last, the curves
are labeled with the values of the resonance energy, $E_\mathrm{R}$ in MeV, and the resonance
strength, $\left(\omega\gamma\right)_\mathrm{R}$ in meV. The hatched region shows the range of
the convective timescales. In this plot, the density is $\rho=2\times10^9$~g~cm$^{-3}$ and the carbon
mass fraction is 0.5. Note that a lower carbon abundance would imply a longer nuclear timescale.
}
\label{fig2}
\end{figure}

We can gain some insight into the impact of a \emph{ghost resonance} on the physics of SNIa by
analyzing the different phases involved in the process of explosive carbon burning:
\begin{itemize}
\item At first, carbon competes with neutrino emission, until the temperature is
high enough to ensure that nuclear heating cannot be balanced by neutrino cooling (ignition condition).
In this work, we define the ignition curve as the local $\rho$-$T$ conditions for which the
nuclear timescale equals the neutrino cooling timescale (see Fig.~\ref{fig2} for the definition
of the nuclear timescale).
Typically, this occurs at $T\sim2\times10^8$~K, just at the border of the temperature range where
the resonance
begins to noticeably increase the carbon fusion rate. The evolution of the white dwarf during this
phase determines the place where carbon ignites: either at the center or close to the white dwarf
surface, the latter giving rise to an accretion-induced collapse of the white dwarf rather
than a supernova event. 
\item When the neutrino cooling is negligible, relative to the nuclear energy release,
convection becomes the dominant energy-transport
mechanism. An adiabatic thermal profile is imprinted in the
white dwarf as long as the convective timescale $\sim10 - 100$~s remains below
the nuclear timescale. If the carbon fusion rate is just given by the
non-resonant contribution, the maximum temperature at which convection is able to remove the
excess nuclear heat is $\sim7 - 9\times10^8$~K. When neither neutrinos nor
convection
can equilibrate the nuclear heating, a flame starts to propagate through the white dwarf. 
Here, we define the dynamical curves as the local $\rho$-$T$ conditions for which the nuclear
timescale equals the (global) convective timescale. In general, the convective timescale is
difficult to determine and may change over time. However, for the purposes of studying the effects
of a LER, it is enough to use an approximate, constant value, which we have taken to be
$t_\mathrm{conv}=10$~s.\footnote{In the literature, the term ignition is
often used to describe the conditions (density, temperature,
chemical composition, and geometry of the runaway kernels) when the white dwarf evolution becomes
dynamical. This should not be confused with our use of the term ignition, referred to in the
ignition curve, which takes place tens, or even hundreds, of years before the actual explosion.}

\end{itemize}
Once a combustion front is present in the white dwarf, it propagates in a way
that is nearly independent of the details of the nuclear reactions, as long as
they are brought to completion. The front can propagate either as a subsonic
flame or as a supersonic detonation.
\begin{itemize}
\item The (microscopic) thermal structure of a subsonic flame
(deflagration) is determined by heat diffusion from hot ashes to cool fuel until the temperature
of the last one is as high as $2 - 5\times10^9$~K (depending on fuel density). At higher
temperatures, the flux of heat released by nuclear reactions is higher than that of the heat diffusion. 
\item Detonations are supersonic combustion waves where ignition is triggered by an initial
temperature jump due to a shock front. In the conditions of thermonuclear supernovae, this jump
raises the fuel temperature well above $\sim2\times10^9$~K. Later on, the heat released by
thermonuclear reactions determines the thermal profile.
\end{itemize}
As can be seen in Fig.~\ref{fig1}, the range of temperatures for which nuclear reactions control the
thermal evolution of thermonuclear combustion fronts in SNIa is above the range of temperatures at
which a \emph{ghost resonance} significantly affects the rate of the carbon fusion reaction.
Thus, we expect that the largest effect of such a resonance will take place during the carbon
simmering
phase.

In this work, we systematically explore the effects of a LER at different
values of the resonance energy, $E_\mathrm{R}$, and with different resonance strengths,
$\left(\omega\gamma\right)_\mathrm{R}$. Moreover, motivated by the experimental finding of an 
alpha-to-proton yield ratio, $\alpha$/p, substantially different from one at the lowest measured energy
(S07), we further
study the effects of varying this ratio. As we demonstrate in Section~\ref{neut}, varying 
$\alpha$/p has an interesting effect on the level of neutronization attained during the carbon
simmering phase. In Table~\ref{tab1}, we summarize the combinations of the resonance
parameters we have considered. Model CF88 is a reference model in which the $^{12}$C+$^{12}$C rate
is given just by the non-resonant CF88 rate. The rate of all the other models is given by
the sum of the CF88 rate and the experimentally known resonance at 2.14~MeV plus a \emph{ghost
resonance} (Eq.~\ref{eqres}). 

\begin{table}[tb]
\caption{Summary of $^{12}$C+$^{12}$C rates explored in this paper.}\label{tab1}
\centering
\begin{tabular}{lccc} 
\hline\hline             
Model & $E_\mathrm{R}$
& $\left(\omega\gamma\right)_\mathrm{R}$ & $\alpha/\mathrm{p}$ \\
 & (MeV) & (meV) & \\
\hline
CF88 & - & - & - \\
LER-1.5-0.1-1.0 & 1.5 & 0.1 & 1.0 \\
LER-1.5-0.001-1.0 & 1.5 & 0.001 & 1.0 \\
LER-1.5-10.-1.0 & 1.5 & 10. & 1.0 \\
LER-1.3-0.1-1.0 & 1.3 & 0.1 & 1.0 \\
LER-1.7-0.1-1.0 & 1.7 & 0.1 & 1.0 \\
LER-1.5-0.1-5.0 & 1.5 & 0.1 & 5.0 \\
LER-1.5-0.1-0.2 & 1.5 & 0.1 & 0.2 \\
\hline
\end{tabular}
\tablefoot{Parameters of the low-energy resonance: resonance energy, $E_\mathrm{R}$,
resonance strength, $\left(\omega\gamma\right)_\mathrm{R}$, and ratio of $\alpha$ to proton
yields.}
\end{table}

A first quantitative evaluation of the impact of the \emph{ghost resonance} parameters assumed in
Table~\ref{tab1} can be made by comparing the nuclear timescale to relevant timescales of
the white dwarf. Figure~\ref{fig2} shows the nuclear timescale for the different models, as
functions of temperature, compared to typical values of the
convective turnover timescale ($t_\mathrm{conv}\approx10$-100~s). We
see that the experimental resonance at 2.14~MeV with no \emph{ghost resonance}
(S07) has a small effect on the nuclear timescale. On the other hand, a resonance
at $E_\mathrm{R}=1.5$~MeV with a strength 100 times smaller than that measured at 2.14~MeV (i.e.,
with $\left(\omega\gamma\right)_\mathrm{R}=0.001$~meV) can reduce the runaway temperature by
$\sim2\times10^8$~K, if the density is kept fixed at $\rho=2\times10^9$~g~cm$^{-3}$. Increasing the
resonance strength by two orders of magnitude translates into an approximate reduction in the
runaway temperature by $\sim10^8$~K. The same reduction in the runaway temperature can be achieved
by decreasing the resonance energy by $\sim0.2$~MeV. 

Similar conclusions would result from using the thermal timescale instead of the nuclear timescale.
In general, the thermal timescale is shorter than the nuclear timescale by a factor
$\tau_\mathrm{th}/\tau_\mathrm{nuc}\simeq2c_pT/QN_\mathrm{A}Y_{12}$, where $c_p$ is the specific
heat at constant pressure and $Q$ is the effective heat release per carbon fusion reaction,
$Q\simeq9.1$~MeV \citep{ch08}. For instance, at $T=6\times10^8$~K,
$\tau_\mathrm{th}/\tau_\mathrm{nuc}\simeq0.06$ but, owing to the steep slope of the
$\tau_\mathrm{nuc}$ vs. $T$ curves (see Fig.~\ref{fig2}), such a reduction in the timescale only
slightly affects the values of the runaway temperatures.

The utility of Fig.~\ref{fig2} is limited by the use of a fixed density, independent of
temperature and of the parameters adopted for the \emph{ghost resonance}.
In reality, both the temperature and density evolve during the simmering phase and depend on the
assumed $^{12}$C+$^{12}$C rate. In the calculations reported in the next section, we address both dependences.

\section{White dwarf evolution from accretion to explosion}\label{wdevol}

We now discuss the evolution of a non-rotating WD from the beginning of accretion to
explosion.
We have followed the WD with an hydrostatic code until the first crossing of the dynamical curve at
any zone within the WD. The structure so obtained was then fed into a hydrodynamic supernova code,
where the structure evolved until the star was completely disintegrated (maximum density
$\lesssim0.2$~g~cm$^{-3}$). Both codes are one-dimensional and assume spherical symmetry. Strictly
speaking, the onset of the dynamical event takes place when the evolutionary timescale becomes
shorter than the sound crossing timescale, $t_\mathrm{sound}\simeq0.1$~s, which occurs slightly
later
than the crossing of the dynamical curve. However, when the nuclear timescale
is shorter than the convective timescale the runaway becomes local, hence the heat released
before attaining $t_\mathrm{sound}$ remains small relative to the WD binding energy,
and the WD structure does not change
appreciably in the interim.

\subsection{Methods}

Our initial model is the same as in \citet{pie03b}, i.e. a CO WD with $M=0.8$~M$_\odot$, on
which we accrete directly CO-rich matter at a rate of $\dot{M}=5\times
10^{-7}$~M$_\odot$~yr$^{-1}$. 
The hydrostatic evolution is performed using the FRANEC evolutionary code
\citep{chi89} with the same  input physics as in \citet{pie03b}. 
The main difference of our approach concerns the neutrino energy losses, which are computed according to 
\citet{esp02} and \citet{esp03}. Moreover, to properly describe the late part of the  
hydrostatic evolution, composition changes produced by convective mixing are
modeled by adding diffusion terms to the nuclear burning terms in the composition equations. 

The supernova hydrodynamic code used in the present work is the same as in \citet{bad03}, in addition to both 
\citet{bra96} and \citet{bra11}. The present models are based on the delayed-detonation paradigm
\citep{kho91}, with a fixed deflagration-detonation transition density,
$\rho_\mathrm{DDT}=1.6\times10^7$~g~cm$^{-3}$.

\subsection{Results}

\begin{figure}[tb]
\centering
  \resizebox{\hsize}{!}{\includegraphics{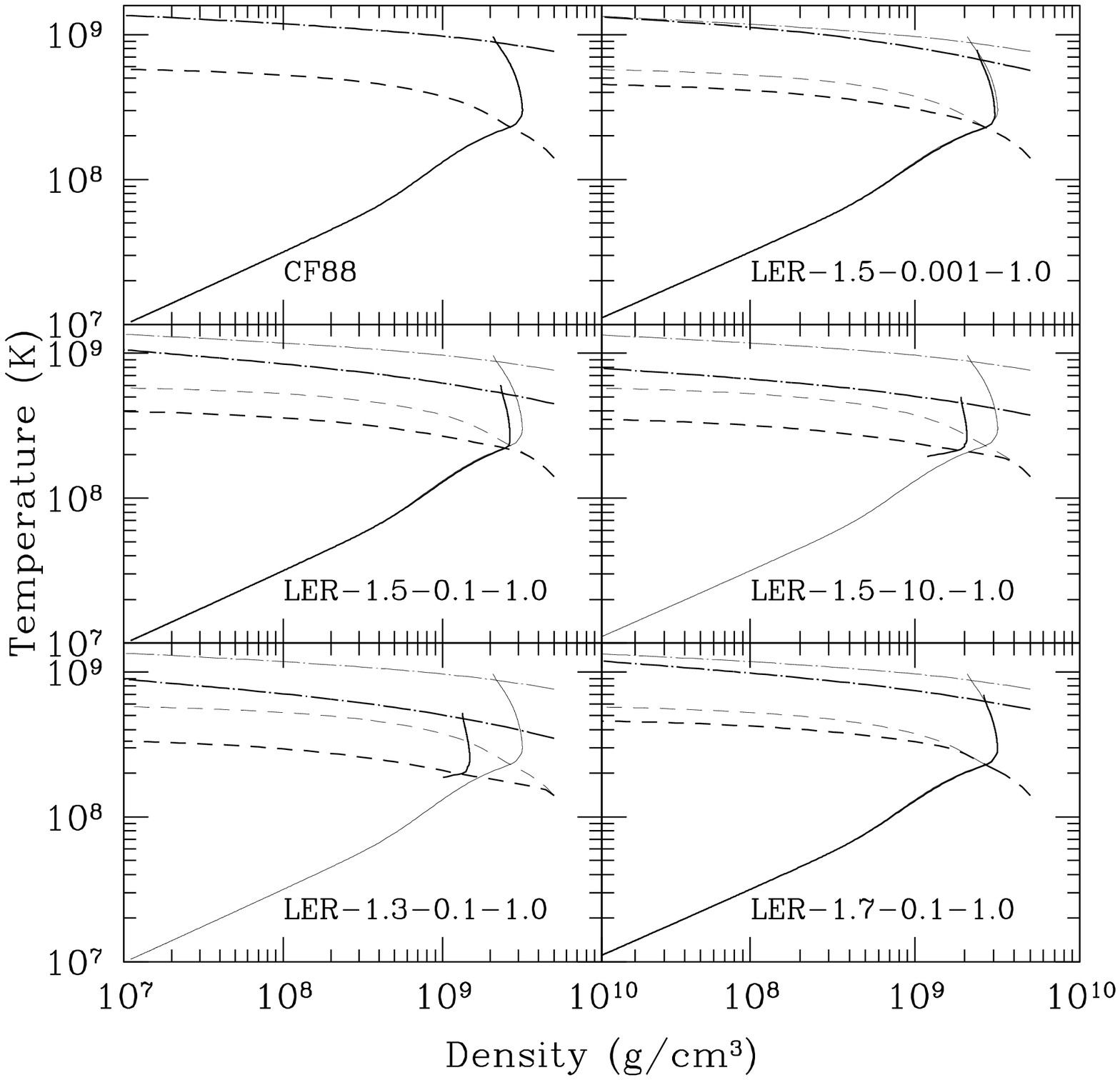}}
\caption{Changes in the physics of the accretion phase owing to a LER: ignition
curves (dashed lines), dynamical curve (dot-dashed lines), and track of central $\rho$ vs. $T$
during the hydrostatic evolution (solid lines). The top left panel shows the results obtained when
the
$^{12}$C+$^{12}$C rate was computed following CF88, the other panels show the results adding a
LER to that rate (see Table 1). For comparison purposes, in all the panels we
have drawn the CF88 curves as thin lines. In the panels belonging to the LER-1.5-10.-1.0
and LER-1.3-0.1-1.0 rates, the track belongs to the evolution of the shell that first crosses the
dynamical curve. 
We assumed a chemical composition made of 50\% $^{12}$C and 50\% $^{16}$O for both curves, while
the hydrostatic $\rho$-$T$ tracks were computed with the realistic chemical composition obtained
from full stellar evolutionary calculations (FRANEC code).
}
\label{fig3}
\end{figure}

\begin{figure}[tb]
\centering
  \resizebox{\hsize}{!}{\includegraphics{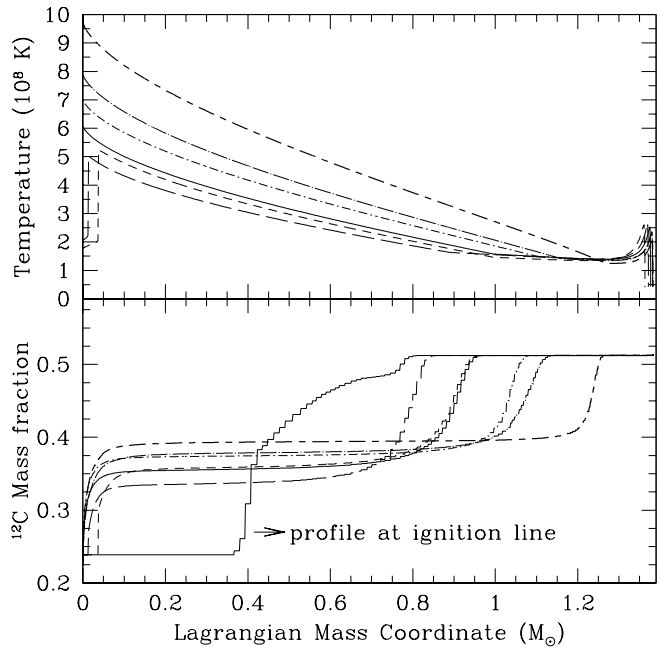}}
\caption{Thermal and chemical profiles at the crossing of the dynamical curves. The line types used
for the models from Table 1 are the same as those used in Fig.~\ref{fig2}. {\bf Top panel:}
Temperature profiles. From top to bottom (at $M\sim0.2$~M$_\odot$) the curves belong to models
CF88, LER-1.5-0.001-1.0, LER-1.7-0.1-1.0, LER-1.5-0.1-1.0, LER-1.3-0.1-1.0, and LER-1.5-10.-1.0.
Note that the last two models run away off-center. {\bf Bottom panel:} $^{12}$C mass fraction
profiles. For comparison purposes, we have plotted with a thin solid line the profile of model CF88
at the crossing of the ignition curve. As can be easily appreciated, the extension of the
convective core is very sensitive to the inclusion of a LER in the
$^{12}$C+$^{12}$C reaction rate.
}
\label{fig4}
\end{figure}

The results of the hydrostatic simulation of the accretion phase up to the crossing of the
dynamical curve are shown in Figs.~\ref{fig3} and \ref{fig4} and Table~\ref{tab2}. Fig.~\ref{fig3}
shows the ignition and dynamical curves obtained using the reference CF88 rate for the
$^{12}$C+$^{12}$C reaction and the changes introduced by the different sets of resonance parameters
that we have explored. The different ignition curves converge towards the CF88 ignition curve
at high densities, whereas the dynamical curves follow the opposite trend, i.e. they diverge
from the CF88 dynamical curve as the density increases. In all cases, as expected, the temperature
defined by these curves, at any given density, decreases when a resonance is included. The effect,
however, depends significantly on the resonance strength and energy. For instance, a resonance
strength of $\left(\omega\gamma\right)_\mathrm{R}=0.001$~meV at $E_\mathrm{R}=1.5$~MeV (model
LER-1.5-0.001-1.0) scarcely affects either the ignition or the dynamical curves. On the other
extreme, the combinations of $\left(\omega\gamma\right)_\mathrm{R}=0.1$~meV at
$E_\mathrm{R}=1.3$~MeV, as well as $\left(\omega\gamma\right)_\mathrm{R}=10.$~meV at
$E_\mathrm{R}=1.5$~MeV (models LER-1.3-0.1-1.0 and LER-1.5-10.-1.0, respectively), result in a
decrease in the temperature of the ignition curve by $\simeq2.5\times10^8$~K at low densities, and
a decrease in the temperature of the dynamical curve by $\simeq4\times10^8$~K at high densities. 

The time elapsed since the beginning of accretion until the supernova explosion is about the same
for all the $^{12}$C+$^{12}$C rates, $t_\mathrm{acc}\simeq1.11$-1.15~Myr, because all the models
need to arrive close to the Chandrasekhar mass before destabilization. Models
LER-1.3-0.1-1.0 and LER-1.5-10.-1.0 are the ones that differ most from the reference model, for which
$t_\mathrm{acc}$ is approximately $40,000$~yr shorter than that of CF88. On the other hand, the
time elapsed between the crossing of the ignition and the dynamical curves
(simmering time, last column of Table~\ref{tab2}) is quite different from model to model. Even for
model LER-1.5-0.1-1.0, whose strength is similar to that of the experimental resonance at 2.14~MeV,
the simmering time is just $\sim13$\% of that of model CF88. 

\begin{table*}[tb]
\caption{White dwarf properties at the crossing of the ignition and dynamical curves.}\label{tab2}
\centering
\begin{tabular}{lllllcccr} 
\hline\hline             
\rule{0mm}{4mm}
 & \multicolumn{3}{c}{ignition curve} & & & & & \\
\rule{0mm}{4mm}
 Model & $T_\mathrm{c}$\tablefootmark{a} & $\rho_\mathrm{c}$\tablefootmark{a} &
$X_\mathrm{c}\left(^{12}\mathrm{C}\right)$\tablefootmark{a} &
$M\left(^{12}\mathrm{C}\right)$\tablefootmark{a} & & & & \\
 & $(10^8~\mathrm{K})$ & $(\mathrm{g}~\mathrm{cm}^{-3})$ & &
$\left(\mathrm{M}_\odot\right)$ & & & & \\
\cline{1-5}
CF88 & 2.658 & $2.85\times10^9$ & 0.2387 & 0.5726 & & & & \\
LER-1.5-0.1-1.0 & 2.256 & $2.56\times10^9$ & 0.2387 & 0.5708 & & & & \\
LER-1.5-0.001-1.0 & 2.357 & $2.84\times10^9$ & 0.2387 & 0.5724 & & & & \\
LER-1.5-10.-1.0 & 2.150 & $2.06\times10^9$ & 0.2387 & 0.5678 & & & & \\
LER-1.3-0.1-1.0 & 1.984 & $1.67\times10^9$ & 0.2388 & 0.5627 & & & & \\
LER-1.7-0.1-1.0 & 2.361 & $2.85\times10^9$ & 0.2387 & 0.5726 & & & & \\
\cline{1-5}
\rule{0mm}{4mm}
 & \multicolumn{3}{c}{dynamical curve} & & & & & \\
\rule{0mm}{4mm}
 Model & $T_\mathrm{c}$\tablefootmark{b} &
$\rho_\mathrm{c}$\tablefootmark{b} & $X_\mathrm{c}\left(^{12}\mathrm{C}\right)$\tablefootmark{b} &
$\Delta M\left(^{12}\mathrm{C}\right)$\tablefootmark{c} & $M_\mathrm{conv}$\tablefootmark{d} &
$M_\mathrm{rw}$\tablefootmark{e} & $R_\mathrm{rw}$\tablefootmark{e} &
$t_\mathrm{simm}$\tablefootmark{f} \\
 & $(10^8~\mathrm{K})$ & $(\mathrm{g}~\mathrm{cm}^{-3})$ & & $\left(\mathrm{M}_\odot\right)$ &
$\left(\mathrm{M}_\odot\right)$ & $\left(\mathrm{M}_\odot\right)$ & (km) & (yr) \\
\hline
CF88 & 9.715 & $2.09\times10^9$ & 0.2237 & 0.0124 & 1.268 & \multicolumn{2}{c}{center} & 534.4 \\
LER-1.5-0.1-1.0 & 6.075 & $2.33\times10^9$ & 0.2332 & 0.0030 & 0.996 & \multicolumn{2}{c}{center} &
71.6 \\
LER-1.5-0.001-1.0 & 7.878 & $2.38\times10^9$ & 0.2272 & 0.0078 & 1.162 & \multicolumn{2}{c}{center}
& 157.9 \\
LER-1.5-10.-1.0 & 5.019\tablefootmark{g} & $1.90\times10^9$\tablefootmark{g} &
0.2708\tablefootmark{g} & 0.0021 & 0.889 & 0.0137 & 149 & 54.2 \\
LER-1.3-0.1-1.0 & 5.240\tablefootmark{h} & $1.34\times10^9$\tablefootmark{h} &
0.3007\tablefootmark{h} & 0.0029 & 0.996 & 0.0382 & 233 & 61.9 \\
LER-1.7-0.1-1.0 & 7.009 & $2.62\times10^9$ & 0.2414 & 0.0050 & 1.102 & \multicolumn{2}{c}{center} &
306.2 \\
\hline
\end{tabular}
\tablefoot{
Initial white dwarf mass: $0.8$~M$_\odot$, accretion rate:
$5\times10^{-7}$~M$_\odot$/yr. 
\tablefoottext{a}{Central temperature, density, and mass fraction and total mass of $^{12}$C at
the crossing of the ignition curve.}
\tablefoottext{b}{Central temperature, density, and mass fraction of $^{12}$C at the crossing of
the dynamical curve.}
\tablefoottext{c}{Total mass of $^{12}$C burned up to the crossing of the dynamical curve.}
\tablefoottext{d}{Size of the convective core.}
\tablefoottext{e}{Location (mass and radius) of the first zone to cross the dynamical curve.}
\tablefoottext{f}{Time elapsed between the crossing of the ignition and dynamical curves, or
simmering time.}
\tablefoottext{g}{The temperature, density and $^{12}$C mass fraction shown belong to the first
zone to cross the dynamical curve. The corresponding values of density and $^{12}$C mass fraction
at the center were: $2.06\times10^9$~g~cm$^{-3}$ and 0.2387, respectively.}
\tablefoottext{h}{The temperature, density and $^{12}$C mass fraction shown belong to the first
zone to cross the dynamical curve. The corresponding values of density and $^{12}$C mass fraction
at the center were: $1.57\times10^9$~g~cm$^{-3}$ and 0.2388, respectively.}
}
\end{table*}

Figure~\ref{fig3} shows the tracks followed by the runaway zone in the $\rho$-$T$ plane. For
models igniting at the center, the track is exactly the same as in CF88 up to the crossing of
the ignition curve, as expected from the discussion in Section~\ref{ler}. In models
LER-1.5-0.001-1.0 and LER-1.7-0.1-1.0, the $\rho$-$T$ tracks are hardly distinguishable from CF88,
even up to the crossing of the dynamical curve. However, as the dynamical curve of the \emph{ghost
resonance} models lies systematically below that of CF88, the dynamical curve is reached at
a different temperature and density than the reference model. On the other hand, model
LER-1.5-0.1-1.0 diverges away from that of 
CF88 after crossing the ignition curve, which results again in both a substantially lower
temperature and a slightly \emph{higher} density at runaway (see Table~\ref{tab2}). 

The most interesting cases are those of models LER-1.5-10.-1.0 and LER-1.3-0.1-1.0, which are
characterized by either a large resonance strength or a low resonance energy. As can be seen in
Table~\ref{tab2}, in both cases the dynamical curve is first reached  
off-center, at radii 149 and 233~km, respectively. The off-center ignition is caused by
the compressional heating timescale of the WD being slightly longer than the thermal diffusion timescale,
so that the WD is not completely isothermal at carbon ignition. A somewhat different
accretion rate would provide off-center ignitions for different combinations of the resonance
parameters. The igniting zones are located in the $\rho$-$T$ plane to the
left of the track of the center of the WD, and reach the dynamical curve at a much lower
temperature \emph{and density} than model CF88. Thus, although the presence of a LER
always results in a smaller temperature at the crossing of the dynamical curve, the
density can be either lower
or higher than in model CF88 depending on the resonance properties. This lack of monotonicity
reflects as well in other details of the outcome of the supernova explosion, as we later discuss.

Table~\ref{tab2} provides additional properties of the WD at runaway, that are worth discussing. The
size of the convective core determines the degree of dissemination of the products of carbon
simmering throughout the WD prior to the SNIa explosion, with consequences for the chemical
stratification of supernova remnants and the interpretation of their X-ray spectra \citep{bad08a}.
In general, the mass of the convective core, $M_\mathrm{conv}$, decreases monotonically with
increasing resonance strength or decreasing resonance energy. For the extreme case, model
LER-1.5-10.-1.0, $M_\mathrm{conv}$ is 30\% lower than for model CF88. 

Another interesting quantity provided in Table~\ref{tab2} is the total mass of $^{12}$C that is
burned during the simmering phase, $\Delta M(^{12}\mathrm{C})$. This mass varies
strongly from model to model, again monotonically decreasing with either
increasing resonance strength or decreasing resonance energy, because of their lower 
temperature at runaway, which implies a smaller thermal content of the WD. The value of $\Delta
M(^{12}\mathrm{C})$ has a significant effect on the amount of neutronization during carbon
simmering, as we discuss in Section~\ref{neut}.

The thermal and $^{12}$C mass fraction profiles through the WD at the dynamical curve are shown in
Fig.~\ref{fig4}, where the range of the convective core can be clearly appreciated. In all the
models, there is a temperature spike close to the surface that is caused by the high accretion rate, but this
spike has no practical consequences for the supernova explosion. The temperature peaks at the
center in all the models apart from LER-1.3-0.1-1.0 and LER-1.5-10.-1.0. In these models, even though
convection extends all
the way down to the very center of the WD, the $^{12}$C+$^{12}$C rate increases so steeply with
temperature that time-dependent, convective mixing is at late times only efficient in the
neighborhood of the runaway zone. 

The chemical profile of $^{12}$C at the crossing of the ignition curve, shown as a thin line in the
bottom panel of Fig.~\ref{fig4}, reflects the previous history of the WD. Above 0.8~M$_\odot$, the
carbon mass fraction is 0.5, which is the value we adopted for the accreting carbon mass fraction, while the
central $\sim0.4$~M$_\odot$ have a homogeneous composition with a smaller carbon mass fraction,
$X_\mathrm{c}(^{12}\mathrm{C})=0.24$, as results from hydrostatic He-burning. 
At runaway, the chemical profile is the result of the competition between convective mixing and
carbon burning. When the dynamical curve is approached, burning consumes carbon more efficiently than
convection can supply fresh fuel, and there is a drop in the carbon mass fraction at the
burning place. The main impact of a LER on the chemical profile is a lower abundance of carbon
in the convective core, owing to the smaller size of the convective region than for model
CF88. On the other hand, models igniting carbon off-center maintain the original $^{12}$C
abundance at the center, because there is neither mixing nor burning there.
However, the differences in the carbon mass fractions and profiles that we have found do not
have a significant impact on the SNIa outcome.

There is currently no successful theoretical model able to predict the
properties of SNIa explosions univocally, starting from a given presupernova
structure. In other words, all current SNIa models rely on a more or less
large number of free parameters, either related to the ignition configuration (usually in two and
three dimensional models),
or to the nuclear front propagation (in models of any dimensionality), or to both of
them. It turns out that the
explosion properties are mainly sensitive to these free, uncertain, parameters rather than
to the details of the presupernova structure. In our study, the adoption of the
delayed-detonation paradigm with a fixed transition density, $\rho_\mathrm{DDT}$, and the use of a
one-dimensional model implies that all the details of the presupernova are lost apart from the central
density, and the chemical composition profile, whose most relevant byproduct is the profile of the
neutron excess. 
As a result, the kinetic energy of the computed SNIa (Table~\ref{tab3}) changes by less than
$\sim5$\% among our set of models. On the other hand, the mass of radioactive
$^{56}$Ni synthesized in the explosion varies by as much as $\sim17$\%. Rather surprisingly, the
largest difference in the $^{56}$Ni mass ejected, $\Delta M(^{56}\mathrm{Ni})\sim0.06$~M$_\odot$,
belongs to two models that have quite similar properties during the accretion phase: 
LER-1.3-0.1-1.0 and LER-1.5-10.-1.0. In any case, the synthesized masses of $^{56}$Ni are well
within the values
usually accepted for normal SNIa, so that the presence of a LER would not change our
basic picture of the formation of the light curve of these standard candles.

\begin{table*}[tb]
\caption{Results of the hydrodynamic explosion models plus post-processing
nucleosynthesis}\label{tab3}
\centering
\begin{tabular}{lcccccc} 
\hline\hline 
Model & $K_{51}$\tablefootmark{a} &
$M(^{56}\mathrm{Ni})$ & $M\left(\mathrm{C}\right)$ & $M\left(\mathrm{O}\right)$ &
$M\left(\mathrm{Mg}\right)$ & $M\left(\mathrm{Si}\right)$ \\
 & &
$\left(\mathrm{M}_\odot\right)$ & $\left(\mathrm{M}_\odot\right)$ & $\left(\mathrm{M}_\odot\right)$
& $\left(\mathrm{M}_\odot\right)$ & $\left(\mathrm{M}_\odot\right)$ \\
\hline
CF88 & 0.912 & 0.379 & $6.2\times10^{-3}$ & 0.242 & 0.045 & 0.321 \\
LER-1.5-0.1-1.0 & 0.959 & 0.371 & $5.3\times10^{-3}$ & 0.201 & 0.045 & 0.341 \\
LER-1.5-0.001-1.0 & 0.933 & 0.374 & $5.7\times10^{-3}$ & 0.229 & 0.049 & 0.332 \\
LER-1.5-10.-1.0 & 0.957 & 0.351 & $1.8\times10^{-3}$ & 0.202 & 0.046 & 0.357 \\
LER-1.3-0.1-1.0 & 0.949 & 0.411 & $1.6\times10^{-3}$ & 0.189 & 0.042 & 0.329 \\
LER-1.7-0.1-1.0 & 0.943 & 0.369 & $5.7\times10^{-3}$ & 0.219 & 0.045 & 0.344 \\
\hline
\end{tabular}
\tablefoot{Each model was computed as a delayed detonation with transition density
$\rho_\mathrm{DDT}=1.6\times10^7$~g~cm$^{-3}$.
\tablefoottext{a}{Kinetic energy in $10^{51}$~erg.}
}
\end{table*}

In Table~\ref{tab3}, we report the yields of some elements that might be affected by a change in
the rate of carbon burning: carbon, oxygen, magnesium, and silicon. We note, in particular,
that oxygen is destroyed by the flame but partially
regenerated in the outer shells of the supernova by explosive carbon burning.
The most interesting characteristic is the monotonic dependence of the yield of carbon on resonance strength and
energy. Although, in general, the variation in the carbon yield with respect to model CF88 is quite
modest, it is largest in the two models igniting off-center (LER-1.3-0.1-1.0 and LER-1.5-10.-1.0).
However, the reason for the smaller carbon yield of these two models does not rely on the
off-center location of the igniting spot. We have checked it by computing a {\sl mixed model} in
which the pre-supernova structure was that belonging to model LER-1.5-10.-1.0, but the supernova
(and the nucleosynthesis) was computed with the CF88 rate, the result being that the yield of
carbon amounted to $5.4\times10^{-3}$~M$_\odot$, i.e. close to the value obtained for the CF88
model. Hence, the reason for the small yield of carbon obtained in the two models with the
strongest and with the lowest energy resonance has to be found in the higher $^{12}$C+$^{12}$C
rate in the outer layers of the star, where carbon burning is incomplete. A
higher carbon fusion rate implies a smaller flame width \citep{tim92}, which is crucial for
completing the carbon burning process when the flame width becomes comparable to the WD radius
\citep{dom11}.

\begin{figure*}[tb]
\centering
  \includegraphics[width=8.8 cm]{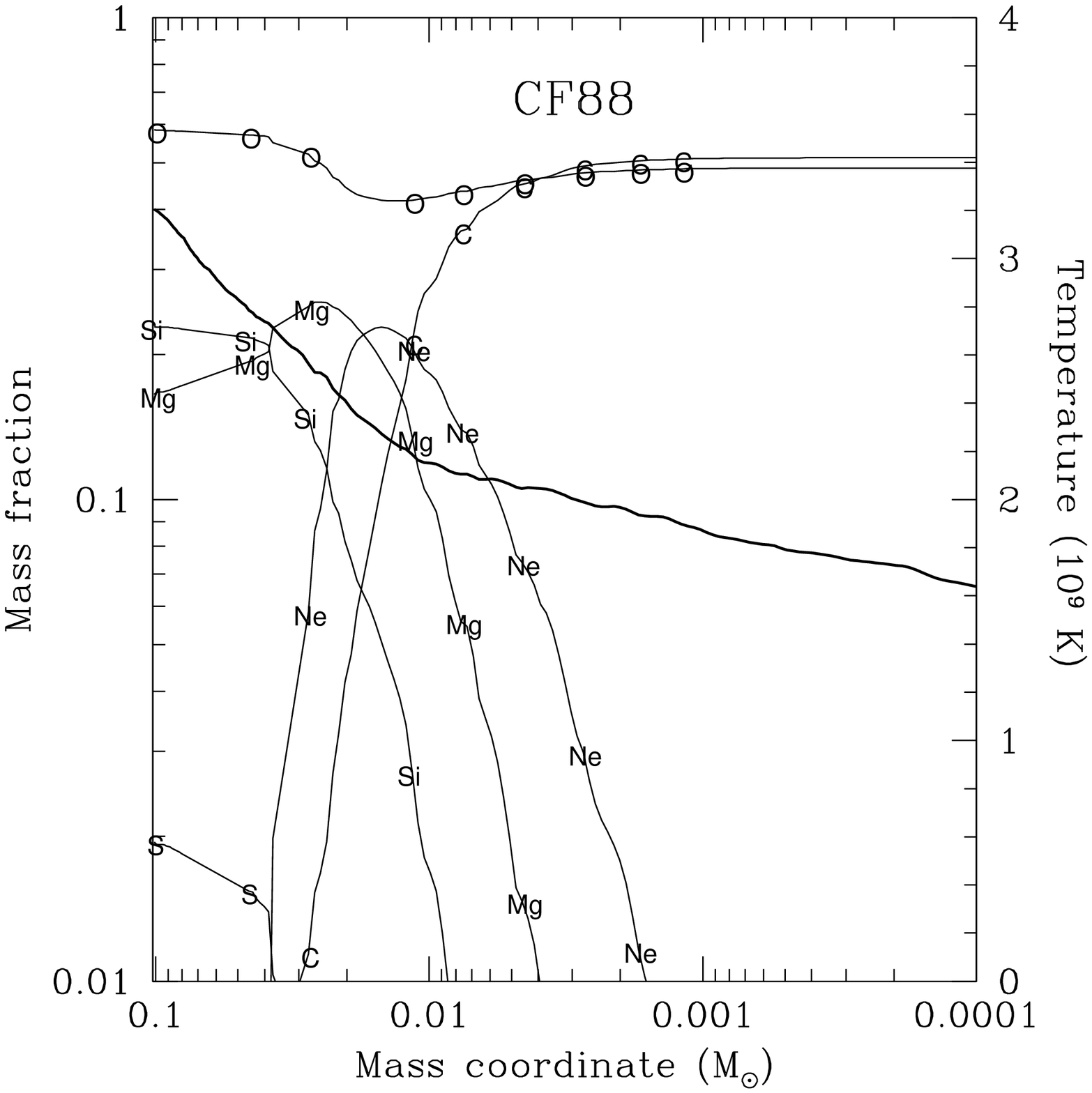}
  \includegraphics[width=8.8 cm]{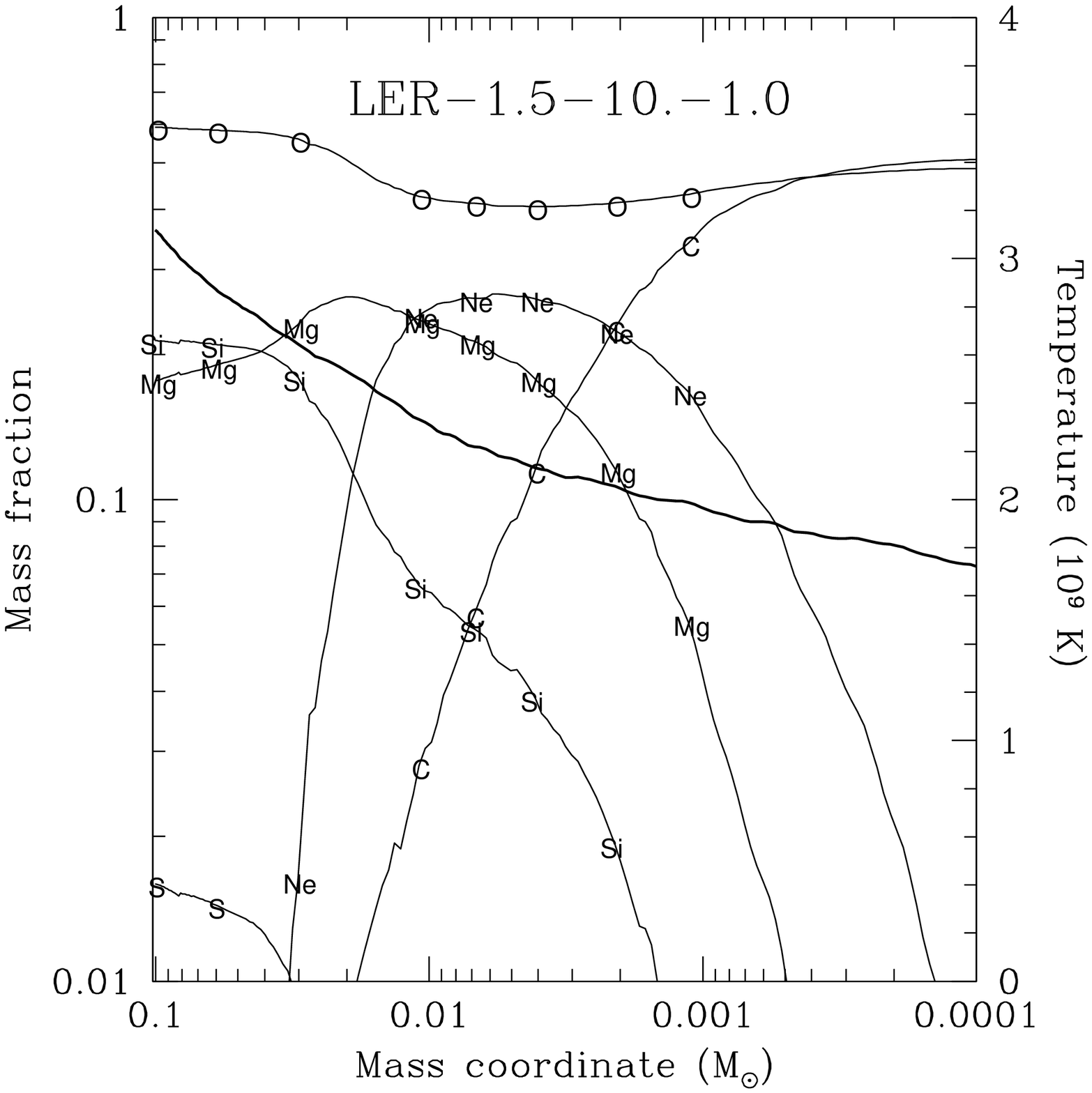}
\caption{Chemical profile in the outer layers of models CF88 and LER-1.5-10.-1.0. The mass
coordinate
measures the mass within the white dwarf surface, i.e. the surface is at the right end of
the horizontal axis. The thicker unlabeled line shows the maximum temperature achieved in each
layer during the explosion (right axis). 
}
\label{fig5}
\end{figure*}

Figure~\ref{fig5} shows the chemical profiles in the outer layers of models CF88 and
LER-1.5-10.-1.0 (note that the mass coordinate in this figure starts at the surface of the ejecta
and increases radially inwards, i.e. to the left of the plot). While the chemical profile is quite similar
in both models within the inner $\sim0.03$~M$_\odot$, carbon is burned more efficiently in the
outermost $\sim0.01$~M$_\odot$ of model LER-1.5-10.-1.0, producing as well a higher yield of
magnesium, neon, and silicon. 
Unfortunately, the yields of the elements most
sensitive to the $^{12}$C+$^{12}$C reaction rate are small and are located in the lowest
density regions of the ejecta, which implies that they can only be detected at a significant
level during the few days following the supernova explosion. 

\section{Physical conditions at thermal runaway}\label{runaway}

The structure of massive non-rotating white dwarfs at carbon runaway is
determined by a few basic properties of the star: 1) its central density,
$\rho_\mathrm{c}$; 2) the electron mole number, $Y_\mathrm{e}$, or the neutron
excess, $\eta=1 - 2Y_\mathrm{e}$; and 3) the number and distribution of hot spots
at which the temperature first runs away. We have discussed in the previous section the dependence
of the first factor, the central density at runaway, on the presence of a LER. The
second factor, the electron mole number, is determined by the weak interactions taking
place during carbon simmering and by convective mixing. Even if we use a time-dependent mixing
scheme coupled to the nuclear burning, the adopted nuclear network is not complete enough to
describe the evolution
of $Y_\mathrm{e}$ with the required precision. In addition, as we now demonstrate, the final
value of $Y_\mathrm{e}$ depends critically on $\alpha$/p. The third factor, namely
the number and distribution of runaway centers, also known as hot spots or igniting bubbles, is
intrinsically multidimensional, so that it cannot be determined with our one-dimensional
hydrostatic code. Hence,
we consider alternative ways of addressing both the dependence of the neutronization during
carbon simmering \citep{ch08,pir08b} and the number and distribution of hot spots, on the presence
of a LER in the $^{12}$C+$^{12}$C reaction \citep{woo04}.

We note that there is another potential effect of the presupernova structure on the
SNIa outcome that we have not accounted for: the possible dependence of the
deflagration-detonation transition density on the WD chemical profile. Although this dependence 
has already been addressed elsewhere \citep{bra10,jac10}, it is somewhat speculative and we prefer
not to mix it with the already hypothetical character of our \emph{ghost resonance}.

\subsection{Neutronization during carbon simmering}\label{neut}

The progenitor metallicity sets a ceiling on the fraction of the incinerated
mass that can be in the form of radioactive $^{56}$Ni, and thus influences the
supernova luminosity, with profound implications for supernova cosmology.
\citet{tim03} proposed a linear relationship between the mass of
$^{56}$Ni and metallicity: $M(^{56}\mathrm{Ni})\propto1 - 0.057Z/Z_\odot$.
Later, \citet{ch08} argued that even a zero metallicity progenitor is
subject to neutronization through electron captures during the time that
elapses between the crossing of the carbon ignition curve and the onset of the
dynamical event, i.e. until the timescale becomes on the order of 1-10~s. The neutronization
during the simmering phase would
affect mainly SNIa from low metallicity progenitors, and the relationship
between $M(^{56}\mathrm{Ni})$ and $Z$ would change slightly, to become
$M(^{56}\mathrm{Ni})\propto0.965 - 0.057Z/Z_\odot$. Existing methods to determine
the progenitor metallicity from the supernova properties depend on the degree of neutronization
during carbon simmering \citep{len00,tau08,bad08a}.

The final neutron excess at runaway is a result of several chains of reactions taking place during
carbon simmering, all of them primed by the $^{12}$C+$^{12}$C reaction. The main reaction chain is
activated by the release of a proton by the carbon fusion reaction 
\begin{eqnarray*}
 ^{12}\mathrm{C} + ^{12}\mathrm{C} &\longrightarrow& ^{23}\mathrm{Na} + \mathrm{p}
\\
 \mathrm{p} + ^{12}\mathrm{C} &\longrightarrow& ^{13}\mathrm{N} \\
 \mathrm{e}^- + ^{13}\mathrm{N} &\longrightarrow& ^{13}\mathrm{C} + \nu_\mathrm{e}
\end{eqnarray*}
\noindent with the result that there is one electron capture for each proton
delivered in the $^{12}\mathrm{C} + ^{12}\mathrm{C}$ reaction. This sequence of reactions is fast
enough relative to the fusion of $^{12}$C to ensure that every time a proton is released it is
followed by an electron capture onto $^{13}$N\footnote{The timescale of electron captures onto
$^{13}$N at the temperatures and densities of interest is much shorter than the WD sound crossing
timescale, hence this reaction is always active during carbon simmering.}. 

Another significant contributor to the neutronization of matter during carbon simmering is the
electron capture onto $^{23}$Na
\begin{eqnarray*}
 \mathrm{e}^- + ^{23}\mathrm{Na} \longrightarrow ^{23}\mathrm{Ne} + \nu_\mathrm{e}
\end{eqnarray*}
\noindent that is activated also by the proton channel of the $^{12}$C+$^{12}$C reaction.
However, in this case the electron capture timescale is longer than the convective timescale
implying
that, as the temperature rises approaching the runaway, the strong interactions become faster than
these electron captures. A direct consequence is that the abundance of $^{23}$Ne freezes out at
high temperatures.

Assuming the proton yield estimated by CF88, i.e. 0.43 protons per reaction, 
gives 0.14 electron captures for every $^{12}$C nuclei consumed in the above 
chains. At densities in excess of $10^9$~g~cm$^{-3}$, other electron captures
become energetically feasible and the ratio of electron captures per $^{12}$C
nuclei destroyed increases to a typical value of $\sim0.3$ \citep{ch08}.
On the other hand, a proton yield as low as that found by S07 for the 2.14~MeV
resonance can modify substantially the ratio of electron captures per $^{12}$C
nuclei destroyed. For instance, a yield of 0.16 protons per reaction would give 0.055
electron captures for every $^{12}$C nuclei consumed in the above chains, i.e.
approximately a factor of three reduction in the amount of electrons captured for a
given reduction in the carbon mass fraction. 

A reduction in the amount of electron captures activated by the proton channel allows 
other reactions to contribute significantly to the neutronization. The initial presence of
$^{22}$Ne opens up another route for electron
captures activated by the alpha channel of the carbon fusion reaction
\begin{eqnarray*}
 ^{12}\mathrm{C} + ^{12}\mathrm{C} &\longrightarrow& ^{20}\mathrm{Ne} + \alpha \\
 \alpha + ^{22}\mathrm{Ne} &\longrightarrow& ^{25}\mathrm{Mg} + \mathrm{n} \\
 \mathrm{e}^- + ^{25}\mathrm{Mg} &\longrightarrow& ^{25}\mathrm{Na}\,,
\end{eqnarray*}
\noindent which has to be taken into account in the calculations of neutronization during carbon
simmering.

\subsubsection{Impact of a low-energy resonance}\label{tracks}

To evaluate the impact of a \emph{ghost resonance} on the degree of neutronization during
carbon simmering we post-processed, with our nucleosynthetic code, the
central $\rho$-$T$ tracks obtained with the hydrostatic code (Fig.~\ref{fig3}) until they crossed
the dynamic curve. We did not account for convective mixing, because the changes in
the chemical composition we consider are so moderate that the contribution of fresh fuel
would not have a significant effect on the nuclear network.

Table~\ref{tab4} shows the results obtained for our reference model, CF88, and two
different resonance strengths with $E_\mathrm{R}=1.5$~MeV,
$\left(\omega\gamma\right)_\mathrm{R}=0.1$ and 0.001~meV (first three rows). We also computed
models that differ from LER-1.5-0.1-1.0 in terms of the $\alpha$ to proton yield ratios (last two rows). 
The second column gives $\mathrm{d}Y_\mathrm{e}/\mathrm{d}Y_{12}$, which is actually the average,
during carbon simmering, of the ratio of the reduction in the electron mole number to the reduction
in the $^{12}$C molar fraction. The presence of the \emph{ghost resonance}, clearly does not affect
the average number of electron captures per carbon nuclei destroyed unless the $\alpha$ or proton
channels contribute to the carbon fusion rate in significantly different proportions. An increase (decrease) in
$\alpha$/p by a factor five translates into
a $\sim50$\% decrease (increase) of $\mathrm{d}Y_\mathrm{e}/\mathrm{d}Y_{12}$. The third column of
Table~\ref{tab4} gives the average derivative of the mean molar fraction with respect to the
carbon molar fraction, which reflects the progress of the nuclear network to
increasingly bound and heavy nuclei. This quantity depends strongly on neither the
presence of the \emph{ghost resonance} nor $\alpha$/p. The small decrement in
$\mathrm{d}Y_\mathrm{b}/\mathrm{d}Y_{12}$ for $\alpha$/p=0.2 is due to the destruction of a third
$^{12}$C nuclei, for each carbon fusion reaction, by a radiative proton capture, which depends
directly on the strength of the proton channel in the $^{12}\mathrm{C} + ^{12}\mathrm{C}$ reaction.

\begin{table*}[tb]
\caption{Neutronization during carbon simmering for different $^{12}$C+$^{12}$C rates.}\label{tab4}
\centering
\begin{tabular}{lcccc} 
\hline\hline 
Model & $\mathrm{d}Y_\mathrm{e}/\mathrm{d}Y_{12}$\tablefootmark{a} &
$\mathrm{d}Y_\mathrm{b}/\mathrm{d}Y_{12}$\tablefootmark{b} & $-\left(\Delta
Y_{12}\right)_\mathrm{core}$\tablefootmark{c} &
$-\left(\Delta Y_\mathrm{e}\right)_\mathrm{core}$\tablefootmark{d} \\
 & & & (mol/g) & (mol/g) \\
\hline
CF88 & 0.2182 & 0.4430 & $8.15\times10^{-4}$ & $1.78\times10^{-4}$ \\
LER-1.5-0.1-1.0 & 0.2167 & 0.4170 & $2.51\times10^{-4}$ & $0.54\times10^{-4}$ \\
LER-1.5-0.001-1.0 & 0.2119 & 0.4232 & $5.59\times10^{-4}$ & $1.19\times10^{-4}$ \\
LER-1.5-0.1-5.0 & 0.0945 & 0.4517 & $2.51\times10^{-4}$ & $0.24\times10^{-4}$ \\
LER-1.5-0.1-0.2 & 0.3197 & 0.3920 & $2.51\times10^{-4}$ & $0.80\times10^{-4}$ \\
\hline
\end{tabular}
\tablefoot{                                                                               
\tablefoottext{a}{Average derivative of the electron mole number with respect to the carbon molar
fraction.}                                                                              
\tablefoottext{b}{Average derivative of the mean molar fraction with respect to the carbon molar
fraction.}
\tablefoottext{c}{Total decrement in the $^{12}$C molar fraction {\sl within the convective core}.}
\tablefoottext{d}{Total decrement in the electron mole number {\sl within the convective core}.}
}
\end{table*}

\begin{figure*}[tb]
\centering
  \includegraphics[width=8.8cm]{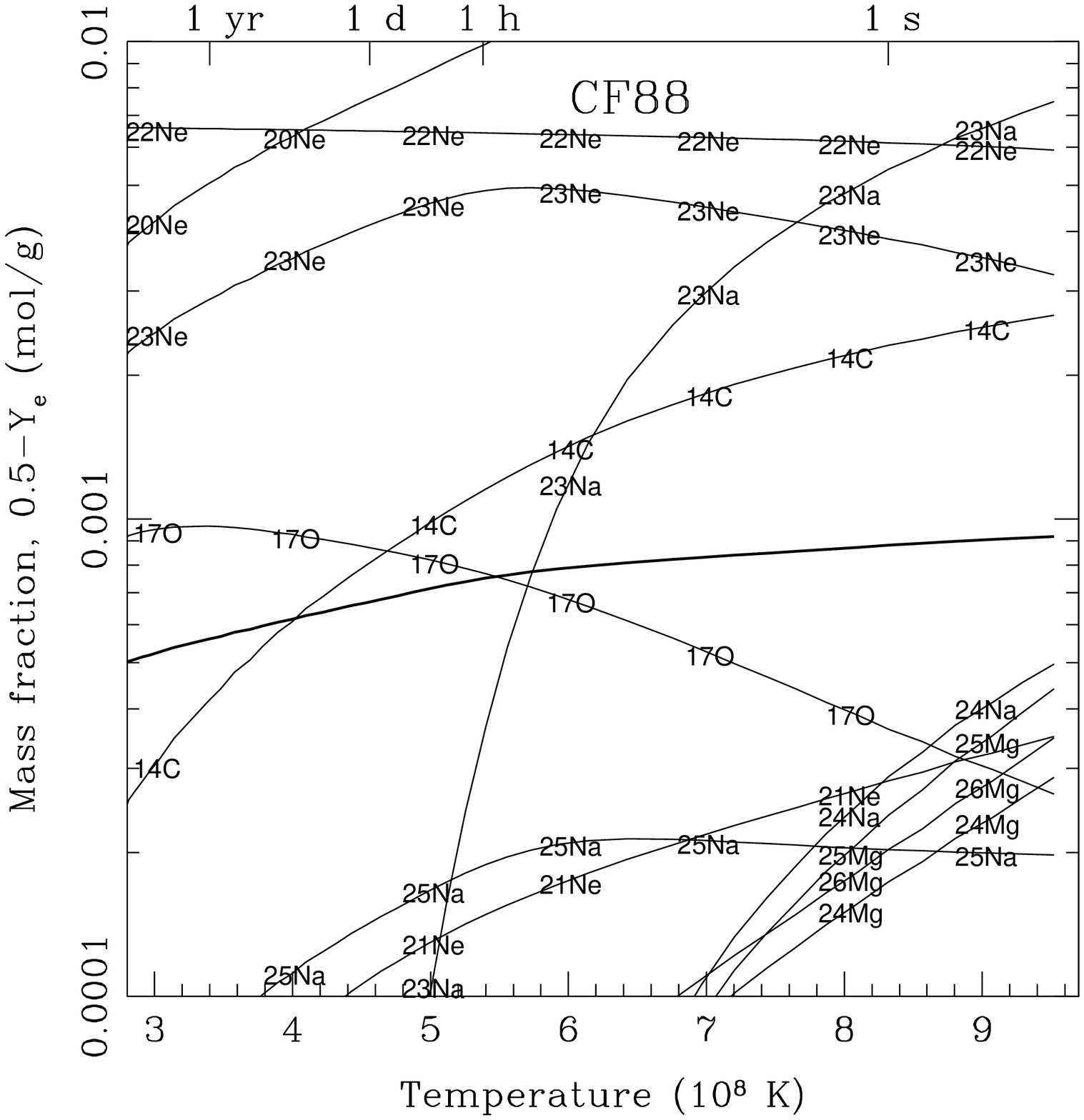}
  \includegraphics[width=8.8cm]{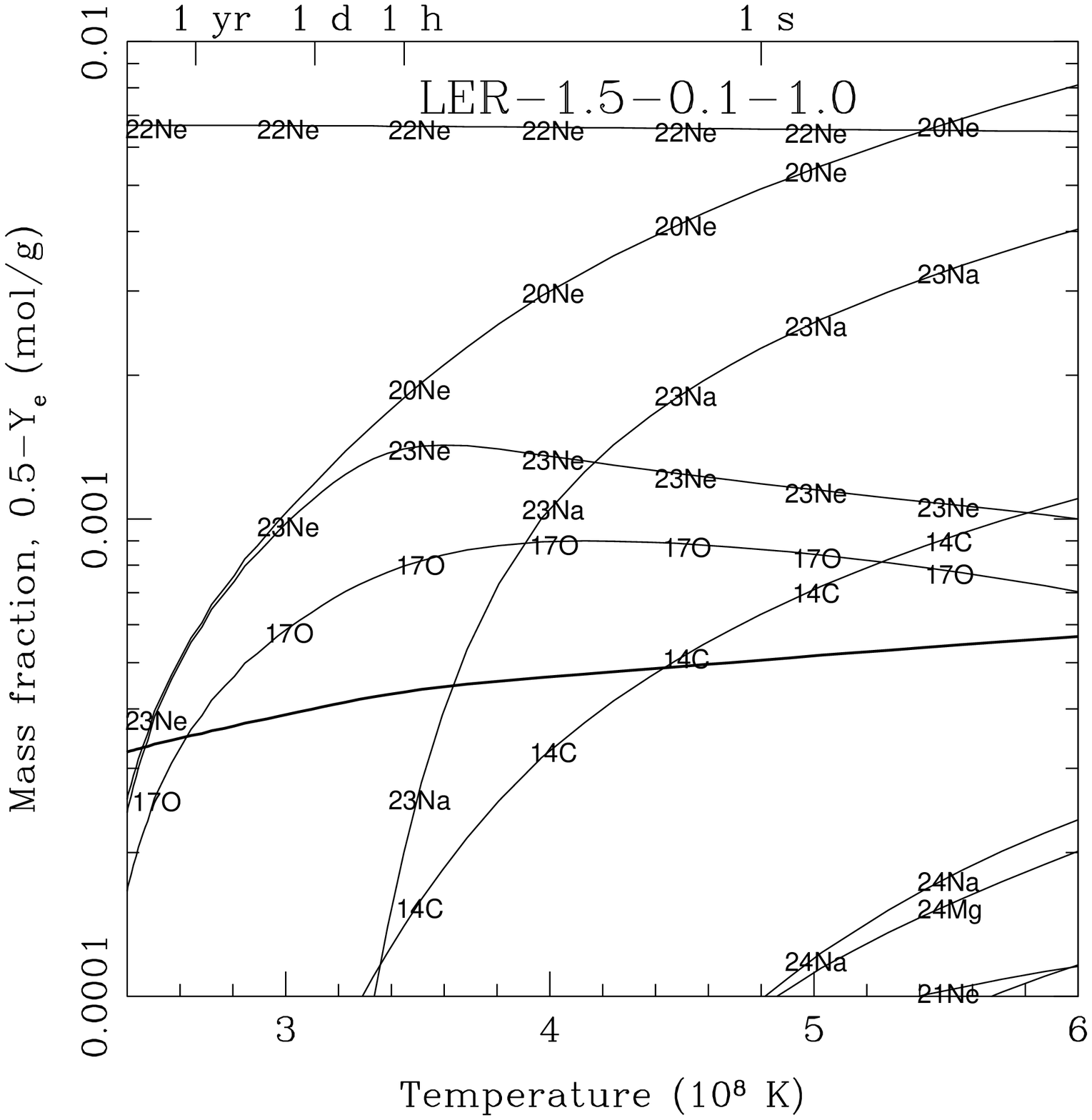} \\
  \includegraphics[width=8.8cm]{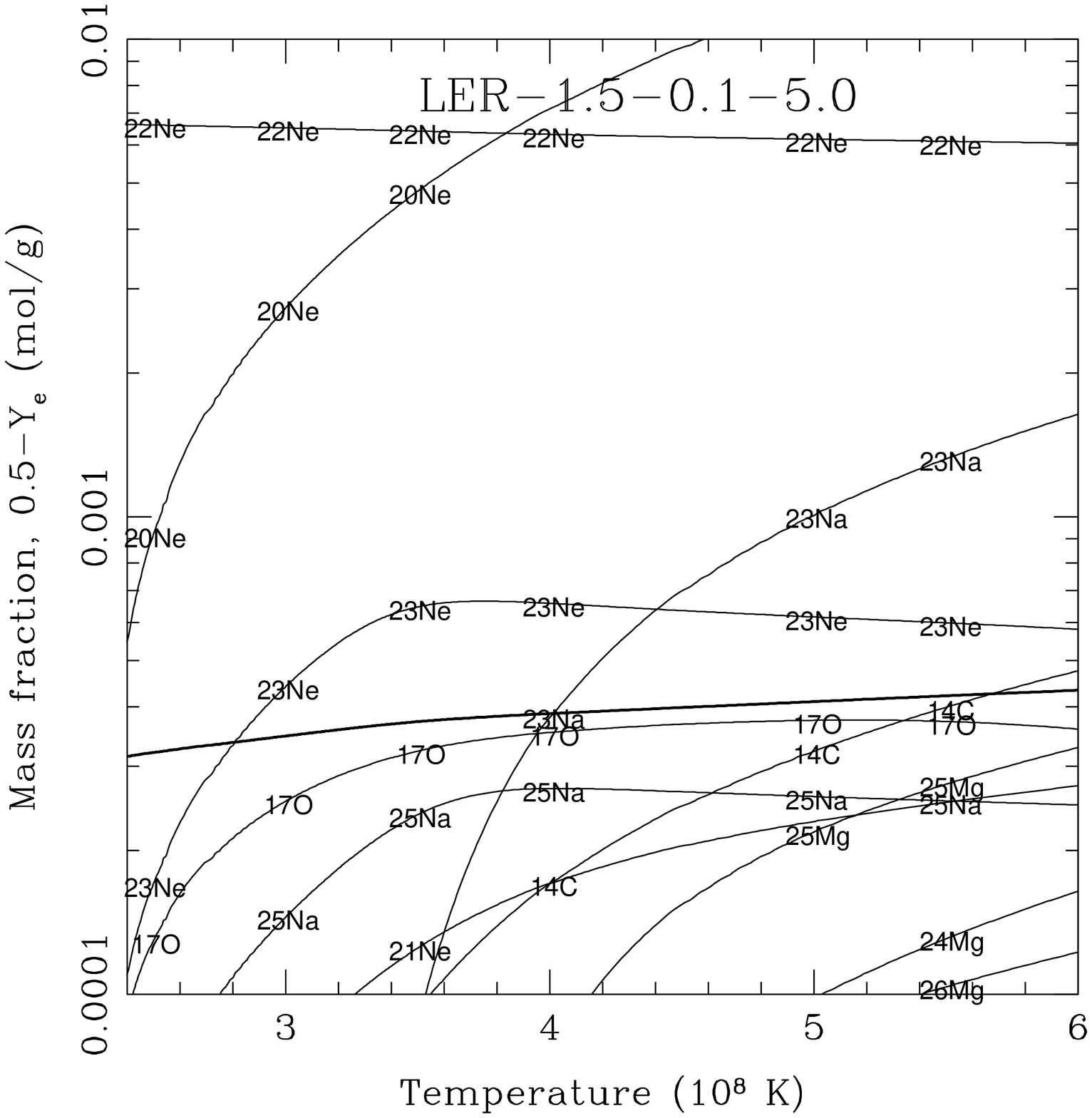}
  \includegraphics[width=8.8cm]{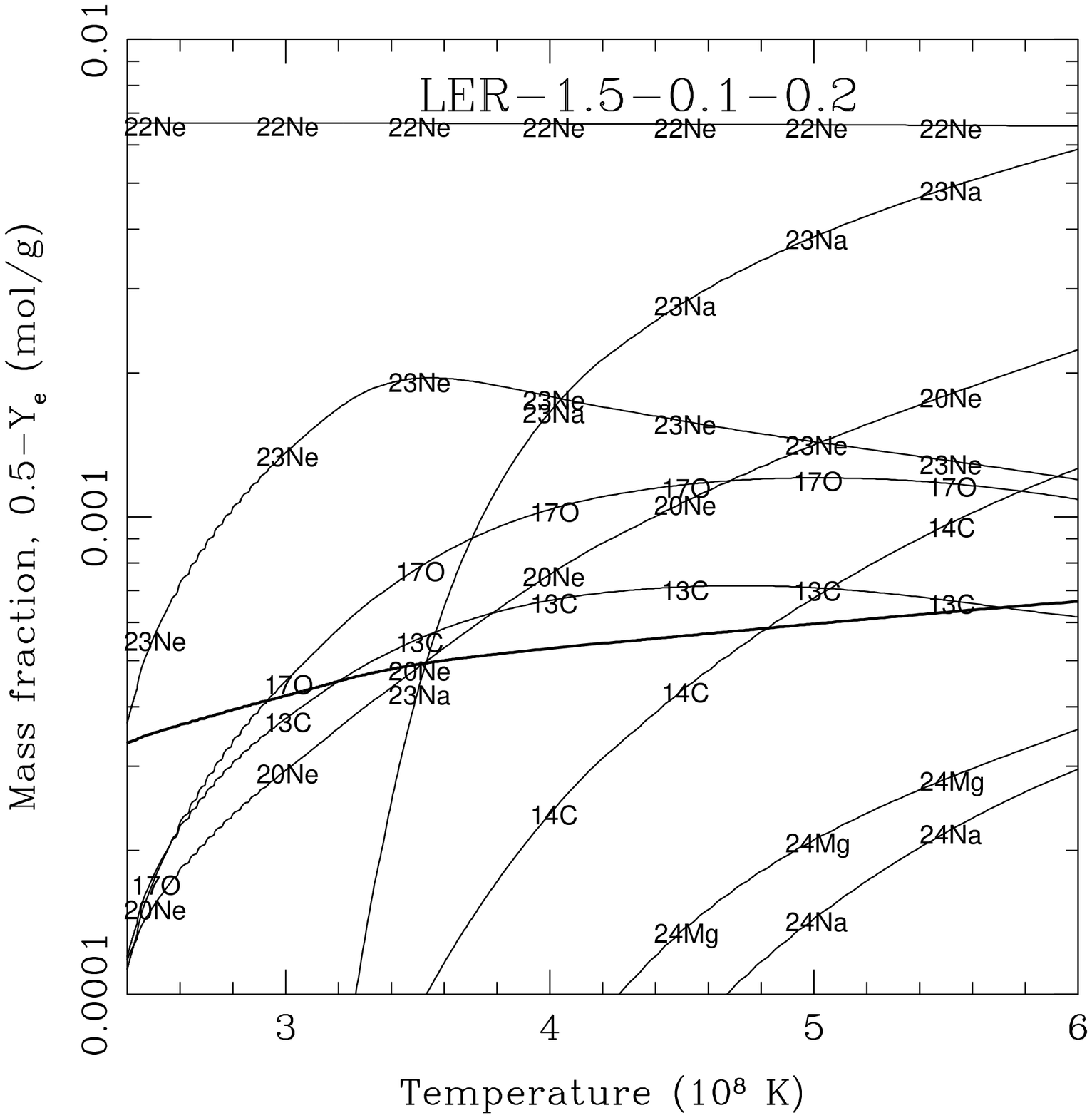}
\caption{Evolution of the chemical composition at the center of the white dwarf as a function of
temperature for different $^{12}$C+$^{12}$C rates: CF88 rate (top left), LER-1.5-0.1-1.0 model
(top right), and $\alpha$/p: LER-1.5-0.1-5.0 with $\alpha$/p=5.0 (bottom
left), and LER-1.5-0.1-0.2 with $\alpha$/p=0.2 (bottom right). To improve the readability, the
upper limit to the vertical axis has been set at a mass fraction of 0.01, hence the curves of
$^{12}$C and $^{16}$O (both with mass fractions $>0.1$) lie out of the plot (the
relative variations in their mass fractions are very tiny within the temperature range shown). The
thick curve gives the electron molar fraction. The
abundance of $^{23}$Ne reflects the amount of electron captures on $^{23}$Na: it grows quickly with increasing temperature at
low temperatures but levels out at $T\sim5.5\times10^8$~K for the CF88 model and
$T\sim3.5\times10^8$~K for
the rest of the models shown. The sum of the abundances of $^{17}$O and $^{14}$C
reflect the electron captures on $^{13}$N: the sum grows steadily but is always smaller than the
abundance of $^{23}$Ne, except at very high temperatures.
The top axes of the plots belonging to models CF88 and LER-1.5-0.1-1.0 are labeled with the time
\emph{to runaway}.
}
\label{fig6}
\end{figure*}

The main conclusion from Table~\ref{tab4} is the definite dependence of the final neutron
excess on the output channel of the carbon fusion reaction. In Fig.~\ref{fig6}, we show details of
the chemical evolution of the reference model, CF88, and the models with $E_\mathrm{R}=1.5$~MeV
and $\left(\omega\gamma\right)_\mathrm{R}=0.1~\mathrm{meV}$ and different $\alpha$/p.
In this figure, the different roles played by the electron captures on
$^{13}$N and $^{23}$Na with respect to neutronization can be appreciated. 

The abundance of $^{23}$Ne directly reflects
the electron captures on $^{23}$Na, it increases steadily at low temperatures but freezes
out as soon as the nuclear timescale becomes smaller than its electron capture timescale. This
condition is achieved at $T\sim5.5\times10^8$~K with the CF88 rate, and at $T\sim3.5\times10^8$~K
with the (low-energy) resonant rate. After peaking at these temperatures, the abundance of
$^{23}$Ne slightly decreases as a consequence of the reaction
\begin{eqnarray*}
 \mathrm{p} + ^{23}\mathrm{Ne} \longrightarrow ^{23}\mathrm{Na} + \mathrm{n}\,.
\end{eqnarray*}
\noindent This last reaction, which leads to a stronger decrease in the $^{23}$Ne abundance in the
models as $\alpha$/p decreases, does not change the electron mole number. Hence, the peak of
the abundance of $^{23}$Ne gives the measure of the contribution of the electron captures
on $^{23}$Na to the global neutronization of matter.

The evolution of the electron captures on $^{13}$N can be monitored by adding the abundances of
$^{14}$C and $^{17}$O. Following each electron capture on $^{13}$N, to give $^{13}$C, this last
nucleus captures any available $\alpha$ particle and unties a chain of reactions 
\begin{eqnarray*}
 \alpha+^{13}\mathrm{C} &\longrightarrow& ^{16}\mathrm{O} + \mathrm{n} \\
 \mathrm{n} + ^{16}\mathrm{O} &\longrightarrow& ^{17}\mathrm{O} + \gamma \\
 \mathrm{n} + ^{17}\mathrm{O} &\longrightarrow& ^{14}\mathrm{C} + \alpha\,. 
\end{eqnarray*}
\noindent Owing to the large abundance of $^{16}$O, the last neutron capture is only significant
after the mass fraction of $^{17}$O has grown to $X(^{17}\mathrm{O})\sim4\times10^{-4}$-$10^{-3}$.
As can be seen by comparing the peak abundance of $^{23}$Ne
with the sum of the abundances of $^{14}$C and $^{17}$O, the largest
contributors to neutronization in all the models are the electron captures onto $^{23}$Na.

Electron captures onto $^{25}$Mg only contribute significantly to
neutronization in model LER-1.5-0.1-5.0, with $\alpha/\mathrm{p}=5$. Even in a model with such
a reduced proton yield, their contribution to neutronization, as measured by the peak abundance of
$^{25}$Na, is less than half the contribution of the electron captures onto $^{23}$Na.

\begin{figure}[tb]
\centering
  \resizebox{\hsize}{!}{\includegraphics{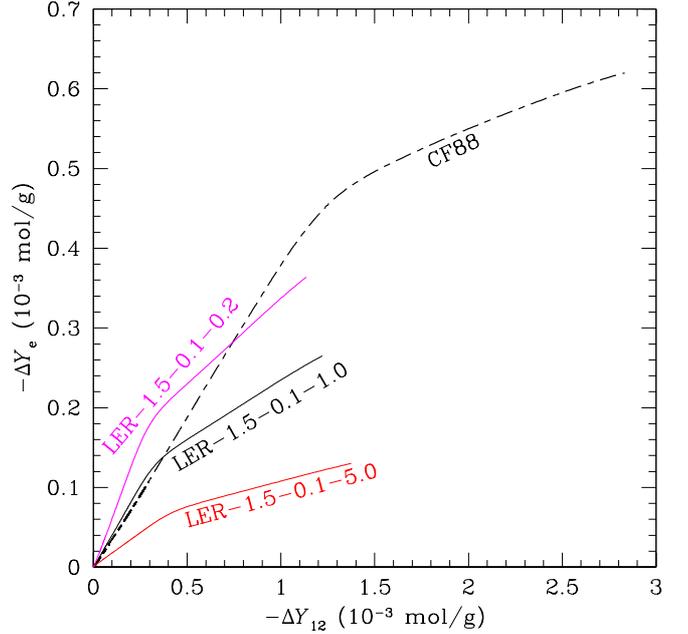}}
\caption{
Evolution of the decrement in the electron mole number vs. the decrement in the carbon
molar fraction for models: CF88, LER-1.5-0.1-1.0, LER-1.5-0.1-5.0, and
LER-1.5-0.1-0.2.
}
\label{fig7}
\end{figure}

Figure~\ref{fig7} shows the evolution of the decrement in the electron mole number as a function
of the decrement in the carbon molar fraction, for the same models as above. Each curve displays
two parts with different slopes. The knee of the curves belongs to
the time at which electron captures on $^{23}$Na freeze-out and, thereafter matter is 
less efficiently neutronized. The plot clearly shows that the models with the same $\alpha$/p 
(CF88 and LER-1.5-0.1-1.0) are characterized by the same rate of neutronization relative to the
rate of consumption of $^{12}$C, while a different weight of proton and $\alpha$ channels implies
different slopes in both parts of the corresponding curves. 
Another interesting feature in Fig.~\ref{fig7} is that the total consumption of $^{12}$C in the
resonant models is far less than in the CF88 model, and the same applies to the
total decrease in the electron mole number.

To evaluate the consequences of any dependence of $-\Delta Y_\mathrm{e}$ on
the resonance properties, we have to take into account the total mass of $^{12}$C
destroyed before reaching the dynamic curve, $\Delta M (^{12}\mathrm{C})$, and the size
of the convective core, $M_\mathrm{conv}$ (Table~\ref{tab1}). The total decrement in the $^{12}$C
molar fraction within the convective core, $-\left(\Delta Y_{12}\right)_\mathrm{core}$ (third
column in Table~\ref{tab4}) is given by
\begin{equation}
 \left(\Delta Y_{12}\right)_\mathrm{core} = -\frac{\Delta M (^{12}\mathrm{C})}{12
M_\mathrm{conv}}\,.
\end{equation}
\noindent The total decrement in the electron mole number within the WD core during carbon
simmering is 
\begin{equation}
 \left(\Delta Y_\mathrm{e}\right)_\mathrm{core} = \frac{\mathrm{d}Y_\mathrm{e}}{\mathrm{d}Y_{12}}
\left(\Delta Y_{12}\right)_\mathrm{core}\,,
\end{equation}
\noindent which is given in the last column of Table~\ref{tab4}. Accounting for all these factors
results in a significant reduction in the neutronization when there is a LER in
the carbon fusion reaction. The largest reduction in $\left(\Delta
Y_\mathrm{e}\right)_\mathrm{core}$ belongs to the model assuming $\alpha/\mathrm{p}=5.0$,
in which case the reduction in the electron mole number in the convective core is only $\sim13$\%
of the reduction obtained with the CF88 model. 

We note that the reduction in the electron mole number within the convective core is
the result of different factors, some of which depend on $\alpha$/p, while others
depend on the resonance energy and strength. However, for $\alpha/\mathrm{p}=5$, even a small
strength of LER (for which $\Delta M (^{12}\mathrm{C})$ and $M_\mathrm{conv}$ were
similar to the CF88 model) would be enough to reduce the neutronization during carbon simmering by
a factor $\lesssim0.5$. 

\subsubsection{Variations in either central density or accretion rate}

\begin{figure}[tb]
\centering
  \resizebox{\hsize}{!}{\includegraphics{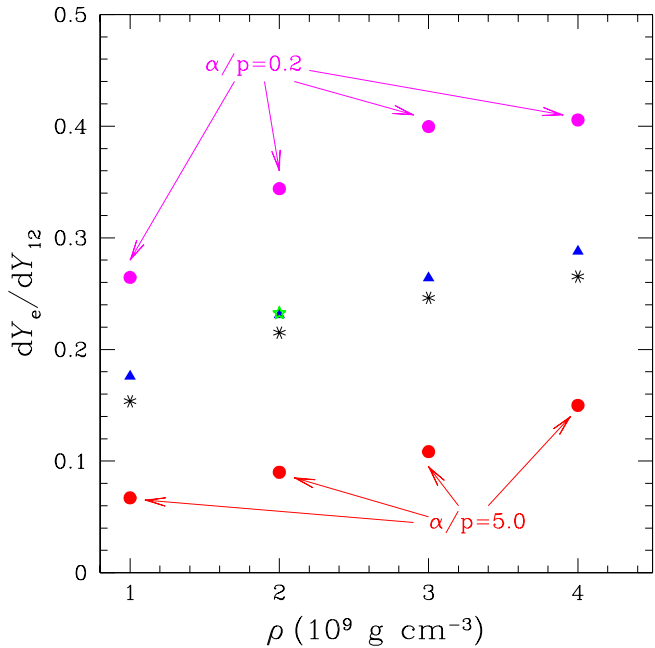}}
\caption{
Average derivative of the electron molar fraction with respect to the $^{12}$C molar
fraction during carbon simmering, as a function of density for different carbon fusion rates: CF88
(black asterisks), LER-1.5-0.1-1.0 (blue
triangles), LER-1.5-0.001-1.0 and LER-1.5-10.-1.0 (green stars), and LER-1.5-0.1-5.0 and
LER-1.5-0.1-0.2 (red and magenta circles, respectively). 
All the models shown in this panel were computed at constant density. 
The models with the extreme resonance strengths were computed only at
$\rho=2\times10^9$~g~cm$^{-3}$.
}
\label{fig8}
\end{figure}

In realistic evolutionary models, the central density at carbon ignition is
determined mainly by the accretion rate. However, 
a variation in $\dot{M}$ also affects the thermal profile along the whole WD,
hence the mass coordinate where carbon runs away \citep[e.g.][]{pie03b}.
While it is impractical to explore with our hydrostatic code the whole space of accretion rates
together with all the resonance parameters, we can gain some indirect insight into the consequences
of different accretion rates by computing the evolution during the simmering phase with our
post-processing code at different constant densities. 
Hence, we repeated the calculations described in Section~\ref{tracks} following,
at constant density, the thermal and chemical evolution between the temperatures of the ignition
and dynamic curves. Our results are summarized in Fig.~\ref{fig8}, where we use four different
densities in the range $10^9$-- $4\times10^9$~g~cm$^{-3}$.

The results show that the
impact of the resonance strength (or even of the existence of a LER) is minimal.
The points belonging to models with $E_\mathrm{R}=1.5$~MeV and resonance strengths
differing by four orders of magnitude are indeed indistinguishable (blue triangles and green stars at
$\rho=2\times10^9$~g~cm$^{-3}$).
In contrast, the most influencing parameter is the ratio of the yields of protons to alphas
in the $^{12}$C+$^{12}$C reaction, whose impact on the variation in the electron molar fraction is
by far higher than that of the density. It is remarkable that the impact of the $\alpha$/p
ratio on the electron captures is about the same at high densities as at $\rho=10^9$~g~cm$^{-3}$,
even though in this case the only electron captures allowed are those onto $^{13}$N. 

\subsection{Runaway bubbles}

The thermal structure of the white dwarf, and thus the distribution of hot
spots, at thermal runaway is set during the last seconds before the crossing of the
dynamical curve, when the nuclear timescale becomes of the same order as
the convective turnover time, $\sim10 - 100$~s \citep[][hereafter WWK04]{woo04}. At this point,
the timescale is so short that the temperature of the central region of the WD is no longer
uniform. Hot spots originate, instead at random locations near the center and are moved by the
convective cells across the turbulent core. During their trip, these bubbles are subject to thermal
stresses of different sign: on the one hand, nuclear reactions release heat efficiently while, on
the other hand, thermal exchange with the background cools them. If the thermal fluctuations
induced by convection are large enough, the heat release by nuclear reactions dominates over
adiabatic cooling and a bubble can run away in a time shorter than the turnover timescale. Although
this process is intrinsically three-dimensional, we can model it following the ideas of WWK04. 

To determine the runaway temperature, $T_\mathrm{rw}$, of the convective bubbles, we use the same
nucleosynthetic
code as before. Now, however, the evolution of the bubble temperature is
integrated together with the nuclear network, and takes into account both the nuclear reactions and
the convective cooling
\begin{equation}
 \frac{\mathrm{d}T}{\mathrm{d}t} = \frac{\dot{\varepsilon}_\mathrm{nuc} -
\dot{\varepsilon}_\nu}{c_\mathrm{p}} + v_\mathrm{conv} \left(\frac{\diff
T}{\diff r}\right)_\mathrm{ad}\,,
\label{eqrw}
\end{equation}
\noindent where $c_\mathrm{p}$ is again the heat capacity at constant pressure,
$\dot{\varepsilon}_\mathrm{nuc}$ is the rate of energy release by nuclear reactions including
electron captures, $\dot{\varepsilon}_\nu$ is the rate of neutrino energy loss (although it is not
relevant near runaway, we have included it for completeness), $v_\mathrm{conv}$ is the convective
velocity, which we take here as a free parameter, and $\left(\diff T/\diff r\right)_\mathrm{ad}$
is the adiabatic thermal gradient. The last term in Eq.~\ref{eqrw} models the loss of heat from
the bubble to the background caused by convection (WWK04). 
Equation~\ref{eqrw} can be easily integrated for a given initial temperature, $T_0$, and bubble
distance to the center, $r_0$, which we take as the radius of the zone that first crosses the
dynamical curve.
As we adopt a constant $v_\mathrm{conv}$, we can easily
calculate the position of the bubble at any time, $t$, as $r=r_0 + v_\mathrm{conv}t$.
Using these approximations, the initial bubble temperature, $T_0=T_\mathrm{rw}$, above which the
integration of Eq.~\ref{eqrw} diverges and the radius, $r_\mathrm{rw}$, at the time of runaway can
be computed for a given central density, $\rho_{0}$, and convective velocity.

Table~\ref{tab5} allows a comparison between the runaway temperature, $T_\mathrm{rw}$, obtained
with a LER and the one achieved with the CF88 rate. The
resonance leads to a lower ignition temperature, from $T_\mathrm{rw}\sim8\times10^8$~K to
$T_\mathrm{rw}\sim4$~--~$6\times10^8$~K. The runaway radius, $r_\mathrm{rw}$, is not as sensitive
to
the presence of the \emph{ghost resonance}, as it stays in a narrow range of
$r_\mathrm{rw}\simeq180$~--~240~km for most models. The runaway radius for model LER-1.3-0.1-1.0
is larger because of the off-center crossing of the dynamical curve.

\begin{table*}[tb]
\caption{Runaway conditions for different $^{12}$C+$^{12}$C rates.}\label{tab5}
\centering
\begin{tabular}{lccccc} 
\hline\hline             
Model & $\rho$\tablefootmark{a} & $X\left(^{12}\mathrm{C}\right)$\tablefootmark{a} &
$T_\mathrm{rw}$\tablefootmark{b} & $r_\mathrm{rw}$\tablefootmark{b} 
& $t_\mathrm{rw}$\tablefootmark{b} \\
 & (g$\cdot$cm$^{-3}$) & & $(10^8~\mathrm{K})$ & (km) & (s) \\
\hline
CF88 & $2.09\times10^9$ & 0.224 & 8.14 & 220 & 2.8 \\
LER-1.5-0.1-1.0 & $2.33\times10^9$ & 0.233 & 4.76 & 177 & 2.2 \\
LER-1.5-0.001-1.0 & $2.38\times10^9$ & 0.227 & 6.03 & 225 & 2.8 \\
LER-1.5-10.-1.0 & $1.90\times10^9$ & 0.271 & 4.26 & 242 & 1.2 \\
LER-1.3-0.1-1.0 & $1.34\times10^9$ & 0.301 & 4.43 & 338 & 1.3 \\
LER-1.7-0.1-1.0 & $2.62\times10^9$ & 0.241 & 5.65 & 184 & 2.3 \\
\hline
\end{tabular}
\tablefoot{In these calculations we assumed a constant convective velocity:
$v_\mathrm{conv}=80$~km/s. 
\tablefoottext{a}{Density and $^{12}$C mass fraction at the point that crosses first the dynamical
curve.}
\tablefoottext{b}{Resulting conditions at runaway: minimum initial temperature of the bubble that
runs away, $T_\mathrm{rw}$, runaway radius, $r_\mathrm{rw}$, and runaway time, $t_\mathrm{rw}$.}
}
\end{table*}

We have explored the sensitivity of the bubble runaway conditions of model LER-1.5-0.1-1.0 to the
assumed physico-chemical parameters: convective velocity, $v_\mathrm{conv}$, density, and
$^{12}$C mass fraction, and present the results in Table~\ref{tab6}. As can be seen, for the
chosen combination of parameters, the runaway temperature remains much lower than that of the CF88
model, and is mostly sensitive to the central density of the WD (we recall that the mass
accretion rates that differ from those we assumed would lead to different central densities at runaway). On
the other hand, a decrease in the convective velocity from 80~km~s$^{-1}$ to 20~km~s$^{-1}$ changes
$T_\mathrm{rw}$ by only $0.2\times10^8$~K. The value of the central density of the WD also affects
substantially the runaway radius, which changes by a factor of two between our extreme densities.
However, it is the convective velocity that has the strongest influence on $r_\mathrm{rw}$. If
$v_\mathrm{conv}$ is as small as 20~km~s$^{-1}$, the bubble runs away just at 90~km from the
center of the WD.

\begin{table}[tb]
\caption{Sensitivity of the runaway conditions of model LER-1.5-0.1-1.0 with respect to
physico-chemical parameters.}\label{tab6}
\centering
\begin{tabular}{cccccrc} 
\hline\hline             
$v_\mathrm{conv}$ & $\rho_\mathrm{c}$ & $X\left(^{12}\mathrm{C}\right)$ &
$T_\mathrm{rw}$ & $r_\mathrm{rw}$ & $t_\mathrm{rw}$ \\
(km/s) & (g$\cdot$cm$^{-3}$) & & $(10^8~\mathrm{K})$ & (km) & (s) \\
\hline
20 & $2.33\times10^9$ & 0.233 & 4.55 & 91 & 4.6 \\
80 & $1.00\times10^9$ & 0.233 & 5.51 & 264 & 3.3 \\
80 & $3.00\times10^9$ & 0.233 & 4.53 & 157 & 2.0 \\
80 & $4.00\times10^9$ & 0.233 & 4.26 & 136 & 1.7 \\
80 & $2.33\times10^9$ & 0.100 & 5.18 & 191 & 2.4 \\
80 & $2.33\times10^9$ & 0.400 & 4.53 & 171 & 2.1 \\
\hline
\end{tabular}
\end{table}

Tables~\ref{tab5} and \ref{tab6} give the time, $t_\mathrm{rw}$ it takes for the bubble to run away
after the temperature reaches $T=T_\mathrm{rw}$. In all cases, the runaway time is short enough to allow for
the survival of the hot bubbles against disruption by the Rayleigh-Taylor instability, which is on
the order of $t_\mathrm{RT}\sim5$~s \citep{iap06}. The survival time of the bubbles against
disruption sets a limit to the maximum distance a bubble can travel from its birthplace before
running away. For instance, for our adopted convective velocity of $v_\mathrm{conv}=80$~km~s$^{-1}$,
this distance is $r_\mathrm{max}\sim400$~km.

\begin{figure}[tb]
\centering
  \resizebox{\hsize}{!}{\includegraphics{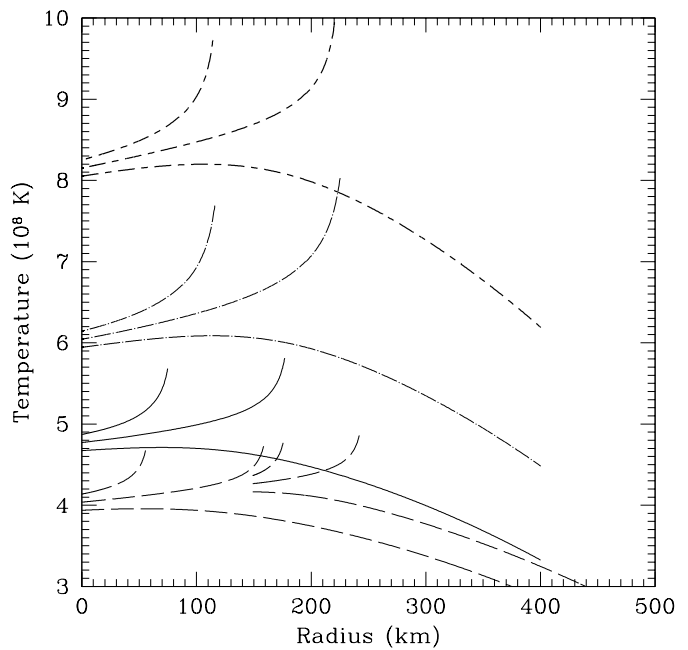}}
\caption{Evolution of the temperature of a burning convective bubble as a function of the distance
to the center of the white dwarf, for different initial temperatures, $T_0$, and prescriptions for
the $^{12}$C+$^{12}$C rate. From top to bottom: CF88 (short-long dashed line), LER-1.5-0.001-1.0
(dot-long dashed line), LER-1.5-0.1-1.0 (solid line), and LER-1.5-10.-1.0 (long-dashed line). For
each reaction rate, we followed the evolution of the bubbles starting at the center, $r=0$,
with different trial initial temperatures, until we found the minimum initial temperature leading
to runaway of the bubble before reaching the end of the convective core.
The distance to the center varies linearly with the elapsed time, $r=r_0+v_\mathrm{conv}t$, where
$v_\mathrm{conv}=80$~km~s$^{-1}$ is the convective velocity. 
To test the impact of the off-center ignition (at $R_\mathrm{rw}\sim149$~km) in model
LER-1.5-10.-1.0, we computed the bubble evolution with the same rate but starting from the
center. The runaway temperature, shown by the thin long-dashed line, is nearly the same as in the
off-center calculation of the same model.
}
\label{fig9}
\end{figure}

Some examples of the temperature evolution of the bubbles are shown in Fig.~\ref{fig9}, for
different initial temperatures, $T_0$, and carbon fusion rates. As can be seen, the radius at
which a bubble runs away depends sensitively on its initial temperature (i.e. on the level of
thermal fluctuations in the convective core). As might be expected, the larger the \emph{excess}
of the initial temperature of a bubble over $T_\mathrm{rw}$, the closer it burns towards its initial
location. 

\subsubsection{Multispot runaway}

Once a bubble runs away, the time available for other sparks to ignite, $\delta t$, is limited by
the time it takes for a conductive flame to cross the whole region where hot spots dwell,
$d\sim2r_\mathrm{max}$. As the conductive flame speed at the densities of interest is
$v_\mathrm{flame}\simeq40$~km~s$^{-1}$ \citep{tim92}, the formation of multiple flame centers is
possible during a time $\sim20$~s. Thus, a multispot runaway scenario is a reasonable assumption
for SNIa, and its properties are an important ingredient for multidimensional SNIa models.

\begin{figure}[tb]
\centering
  \resizebox{\hsize}{!}{\includegraphics{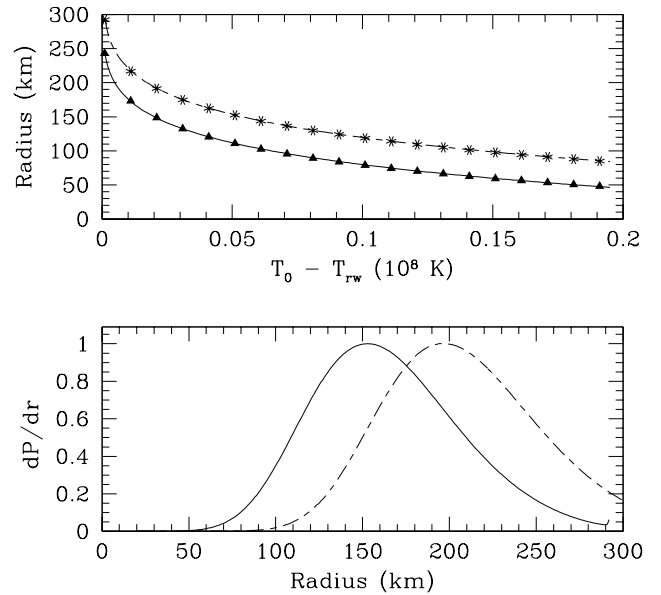}}
\caption{Distribution of runaway radii. {\bf Top.} Runaway radius as a function of the
temperature above the runaway temperature ($T_\mathrm{rw}$, see Table~\ref{tab5}): stars belong to
model CF88, and triangles to model LER-1.5-0.1-1.0. The lines are
quadratic fits to the dependence of runaway radii on temperature: $r=ax^2 + bx + c$, where $r$ is
in km and $x\equiv \ln\left(T_0 - T_\mathrm{rw}\right)$, with both temperatures in $10^8$~K. The
parameters of the polinomial are
$a=-2.87$, $b=-64.08$, and $c=-13.09$ for the CF88 rate, and $a=-2.72$, $b=-60.59$, and $c=-45.21$
for the LER-1.5-0.1-1.0 rate. {\bf Bottom.} Probability distribution function of runaway radii,
in arbitrary units. The probability distribution function has been obtained as the product of the
exponential probability distribution function (EPDF) of temperatures proposed by WWK04 (eq.~38),
$\mathrm{d}P/\mathrm{d}T$, and the inverse of the derivative $\mathrm{d}r/\mathrm{d}T$ computed
from the polinomial fit to $r(T)$. The curves belong to the CF88 rate (short-long dash) and the
LER-1.5-0.1-1.0 rate (solid).
}
\label{fig10}
\end{figure}

Because the induction of local peaks of temperature by convection is a stochastic process, we
have to rely on a statistical approach to the multispot scenario. Indeed, WWK04 studied the
distribution of thermal fluctuations in the convective core of an igniting WD and proposed that
there is an
exponential probability density function (EPDF) of temperatures
\begin{equation}
 \frac{\diff P}{\diff T} \propto - \left(1-f\right)^{\left(T-T_\mathrm{av}\right)/\Delta T}\,,
\label{eqepdf}
\end{equation}
\noindent where $f$ is a parameter correlated to convective mixing efficiency, which following
WWK04 we assume to be $f\simeq0.9$, $T_\mathrm{av}$ is the average temperature,
which we take without loss of generality as the runaway temperature, 
$T_\mathrm{av}=T_\mathrm{rw}$, and $\Delta
T\sim0.1\times10^8$~K or slightly lower (WWK04). To derive a distribution function for
the runaway radii (distance to the center of the WD) of the bubbles, we need the relationship
between $r_\mathrm{rw}$ and the initial temperature excess over $T_\mathrm{av}$. We 
computed, for
both model CF88 and model LER-1.5-0.1-1.0, the temperature evolution of bubbles with an initial
temperature, $T_0$, above
the runaway temperature, from which we found the corresponding runaway radii, and we fit
an empirical relationship between $T_0$ and $r_\mathrm{rw}$ (Fig.~\ref{fig10}). A second-order
polynomial provides a good description of $r_\mathrm{rw}\left(T_0\right)$ and
allows us to find the probability density function of runaway radii
\begin{equation}
 \frac{\diff P}{\diff r_\mathrm{rw}} = \frac{\diff P/\diff T}{\diff r_\mathrm{rw}/\diff T_0}\,.
\end{equation}
\noindent The results are shown in the bottom panel of Fig.~\ref{fig10}. The distributions of
runaway radii for both carbon fusion rates are qualitatively similar in shape and
width, but with the LER the peak of the distribution is $\sim50$~km closer to the
center of the WD. Such a displacement of the runaway radius can have important consequences for
SNIa models such as the gravitationally confined detonation model \citep[see, e.g.,][]{rwh07}.

\section{Conclusions}\label{theend}

We have systematically explored the consequences that a hypothetical low-energy resonance in the
carbon fusion rate would have for the physics and outcome of thermonuclear supernovae. We have
considered resonance energies in the range $E_\mathrm{R}=1.3$~--~1.7~MeV, with strengths limited
by the available experimental cross-section data at
$E_\mathrm{cm}>2.10$~MeV. We have also studied different ratios of the yields of $\alpha$
to
protons, $\alpha/\mathrm{p}=0.2$ -- 5.0. Within these limits, the phase of the supernova that is
most affected by the presence of a LER is carbon simmering. In particular, the degree of
neutronization of the white dwarf prior to the
explosion can be reduced by a large factor even for a LER with a strength
much smaller, say by four-six orders of magnitude, than that measured for the resonance with the
lowest experimental energy ($E_\mathrm{R}=2.14$~MeV, S07). 

The main changes induced by a LER with respect to the model that uses
the standard carbon fusion rate, CF88, are:
\begin{itemize}
 \item The central density of the white dwarf at runaway can increase by $\sim$25--30\%, for a
given mass accretion rate. However, a strong enough resonance gives an off-center ignition, in
which case the density is lower than that obtained with the CF88 rate.

 \item The minimum neutron excess of thermonuclear supernova ejecta, determined by the progenitor
initial metallicity and by several electron capture reactions synchronized with carbon simmering,
is quite sensitive to the relative $\alpha$ to proton yield of the carbon fusion reaction
The presence of a small resonance at 1.5~MeV (1.3~MeV) with a strength
$\gtrsim13$~neV ($\gtrsim0.12$~neV) and an alpha/p yield ratio substantially different from unity
would dominate over the non-resonant contribution at the relevant Gamow energies and, hence, would
have a non-negligible impact on the neutronization of matter during carbon simmering.

 \item The mass of the convective core is also affected by a LER. In general,
the lower the resonance energy and/or the stronger the resonance, the smaller the convective
core. The neutronization produced during carbon simmering erases the imprint of the initial
metallicity of the supernova progenitor, and convection homogenizes the core composition
before the thermal runaway. Hence, the smaller the convective core the higher the mass of the
supernova that bears a record of the initial metallicity.

 \item The time it takes the WD to cover the path from the ignition curve to the dynamic curve
(simmering time) is substantially shorter (by up to a factor ten) if there is a strong LER.

 \item The runaway temperature is strongly affected by the presence of a LER,
which could lower $T_\mathrm{rw}$ by as much as 2~--~$4\times10^8$~K.

 \item In a multispot runaway scenario, the distribution of igniting bubbles changes
quantitatively in the presence of a LER. With such a resonance, even of
moderate strength, bubbles run away closer to the center than predicted by
using the CF88 rate \citep[see also][]{iap10}. On the other hand, if the resonance were strong
enough to provoke an off-center ignition, it would have the opposite effect. 

 \item At the present level of theoretical understanding of the physics of thermonuclear
supernovae, it is difficult to predict the true impact of an increase in the rate
of carbon fusion on the nucleosynthesis and energetics of the explosive event.
Such an study would require a completely parameter-free successful
SNIa model, which is currently not available. Within the one-dimensional delayed-detonation
paradigm, we have found quite small variations in the kinetic energy of the supernova, $K$, mild
variations in the yield of $^{56}$Ni (but still well within the observationally allowed range),
and a ratio of $K$ to the mass of unburned carbon ejected that is quite sensitive to
the existence of a LER.
\end{itemize}

Finally, we briefly discuss some potential (even though speculative) implications
of the distributions of the runaway radii of the ignition bubbles that we have found for the
different models with/without a LER. 
We have shown that both the runaway radius, $r_\mathrm{rw}$, is affected by a LER
(Table~\ref{tab5}), and the runaway bubbles cluster around $r_\mathrm{rw}$
(Fig.~\ref{fig10}). However, although a LER may influence the outcome
of the explosion, it is impossible at present 
to make a more specific prediction. On the other hand, we have shown that $r_\mathrm{rw}$
is a sensitive
function of the central density (Table~\ref{tab6}), with the result that at a higher density the
igniting bubbles would be concentrated toward smaller runaway radii, which can have important
consequences for the subsequent development of the explosion. The effects of the distribution of
hot spots in the
WD at runaway has been explored just recently with the aid of multidimensional
SNIa models. For instance, \citet{sei11} searched for the systematic effects of the central density
on
SNIa luminosity by simulating WD explosions starting from hot bubbles randomly located within a
radius independent of $\rho_\mathrm{c}$. As we have shown, it would be advisable to use a
density-dependent distribution of the bubble positions in these studies, even if the
carbon
fusion rate is that of CF88. 

We conclude that a robust understanding of the links between SNIa properties and their progenitors
will not be attained until the $^{12}$C+$^{12}$C reaction rate is measured at energies
$\sim1.5$~MeV.

\begin{acknowledgements}
This work has been partially supported by the Spanish Ministry for Science and Innovation projects
AYA2008-04211-C02-02, AYA08--1839/ESP, and EUI2009--04170, by European Union FEDER funds, by
the Generalitat de Catalunya, and by the ASI-INAF I/016/07/0. JAE research is funded by the
Comissionat per a Universitats i Recerca of the DIUE of the Generalitat de Catalunya, and by ESF. 
\end{acknowledgements}

\bibliographystyle{aa}
\bibliography{../ebg}

\end{document}